\documentclass[a4paper]{article}
\usepackage{hyperref}
\usepackage{amsmath}
\usepackage{amsthm}
\usepackage{amsfonts}
\usepackage{mathrsfs}
\usepackage{graphicx}
\usepackage{epic,eepic}

\renewcommand\vec{\overrightarrow}

\def\noi{\noindent}

\newtheorem{lemma}{Lemma}
\newtheorem{proposition}{Proposition}
\theoremstyle{definition}
\theoremstyle{remark}

\title{\bf Time-Minimal Control of Dissipative Two-level Quantum Systems:
the  Generic Case}
\author{Bernard~Bonnard,\thanks{B. Bonnard is with the Institut de Math\'ematiques de Bourgogne, UMR CNRS 5584, 9 Avenue
        Alain Savary, BP 47 870 F-21078 DIJON Cedex FRANCE ({\tt bernard.bonnard@u-bourgogne.fr}).}
        ~Monique~Chyba\thanks{M. Chyba is with the University of Hawaii, Department of Mathematics, Honolulu, HI 96822, USA
        ({\tt mchyba@math.hawaii.edu}).}
        ~and~Dominique Sugny\thanks{D. Sugny is with the Institut Carnot de Bourgogne, UMR 5209 CNRS-Universit\'e de Bourgogne,
     9 Av. A. Savary, BP 47 870, F-21078 DIJON Cedex, FRANCE ({\tt
     dominique.sugny@u-bourgogne.fr}).}}

\begin{document}

\maketitle

\begin{abstract}
The objective of this article is to complete preliminary results
from \cite{bonnardsugny,sugnykontz} concerning the time-minimal
control of dissipative two-level quantum systems whose dynamics is
governed by Lindblad equations. The extremal system is described
by a 3D-Hamiltonian depending upon three parameters. We combine
geometric techniques with numerical simulations to deduce the
optimal solutions.
\end{abstract}

\noi\textbf{Keywords.} Time optimal control, conjugate and cut
loci, quantum control\\

\section{Introduction}
In this article, we consider the time-minimal control analysis of
two-level \emph{dissipative} quantum systems whose dynamics is
governed by \emph{Lindblad equation}. More generally, according to
\cite{lindblad}, the dynamics of a finite-dimensional quantum
system in contact with a dissipative environment is described by
the evolution of the \emph{density matrix} $\rho$ given by the
equation
\begin{equation}\label{eqconc0}
i\frac{\partial \rho}{\partial
t}=[H_0+H_1,\rho]+i\mathcal{L}(\rho),
\end{equation}
where $H_0$ is the field-free Hamiltonian of the system, $H_1$
represents the interaction with the control field and
$\mathcal{L}$ the dissipative part of the equation; $[A,B]$ is the
commutator of the operators $A$ and $B$ defined by $[A,B]=AB-BA$.
Equation (\ref{eqconc0}) is written in units such that $\hbar=1$.
The components of the density matrix satisfy the following
equations:
\begin{eqnarray}\label{eqconc00}
\dot{\rho}_{nn}&=&-i[H_0+H_1,\rho]_{nn}-\sum_{k\neq
n}\gamma_{kn}\rho_{nn}+\sum_{k\neq n}\gamma_{nk}\rho_{kk}\nonumber\\
\dot{\rho}_{kn}&=&-i[H_0+H_1,\rho]_{kn}-\Gamma_{kn}\rho_{kn},~k\neq
n
\end{eqnarray}
where $1\leq k\leq N$ and $1\leq n\leq N$ for a $N$-level quantum
system. The parameters $\gamma_{kn}$ describe the population
relaxation from state $k$ to state $n$ whereas $\Gamma_{kn}$ is
the dephasing rate of the transition from state $k$ to state $n$.
Note that not every positive parameter $\gamma_{kn}$ or
$\Gamma_{kn}$ is acceptable from a physical point of view since
the density matrix $\rho$ must satisfy particular properties
(trace conservation, hermitian operator and complete positivity
\cite{altafini,gorini,lindblad}).

Particularizing now to the case $N=2$, we assume that $H_1$ is of
the form
\begin{equation}
H_1=-\mu_x E_x-\mu_y E_y,\nonumber
\end{equation}
where the operators $\mu_x$ and $\mu_y$ are proportional to the
Pauli matrices $\sigma_x$ and $\sigma_y$ in the eigenbasis of
$H_0$. The electric field is the superposition of two linearly
polarized fields $E_x$ and $E_y$ and we assume that these two
fields are in resonance with the Bohr frequency $E_2-E_1$. In the
\emph{RWA approximation}, the time evolution of $\rho(t)$
satisfies the following form of the Lindblad equation
\begin{eqnarray}\label{eqconc1}
i \frac{\partial}{\partial t} \left(
\begin{array}{c}
\rho_{11} \\ \rho_{12} \\ \rho_{21} \\ \rho_{22}
\end{array} \right) =
\left(
\begin{array}{cccc}
-i\gamma_{12} & -E^* & E & i\gamma_{21} \\
-E & -\omega-i\Gamma & 0 &  E \\
E^* & 0 & \omega-i\Gamma & -E^* \\
i\gamma_{12} & E^* & -E & -i\gamma_{21}
\end{array} \right)
\left(
\begin{array}{c}
\rho_{11} \\ \rho_{12} \\ \rho_{21} \\ \rho_{22}
\end{array} \right)
,
\end{eqnarray}
where $E$ is equal to $E=ue^{i\omega t}$ and $u$ is \emph{the
complex Rabi frequency} of the laser field (the real and imaginary
parts of $u$ are the amplitudes of the real fields $E_x$ and $E_y$
up to a multiplicative constant). In Eq. (\ref{eqconc1}), $\omega$
is the difference of energy between the ground and excited states
of the system. In the interaction representation, Eq.
(\ref{eqconc1}) becomes
\begin{eqnarray}\label{eqconc3}
i \frac{\partial}{\partial t} \left(
\begin{array}{c}
\rho_{11} \\ \rho_{12} \\ \rho_{21} \\ \rho_{22}
\end{array} \right) =
\left(
\begin{array}{cccc}
-i\gamma_{12} & -u^* & u & i\gamma_{21} \\
-u & -i\Gamma & 0 &  u \\
u^* & 0 & -i\Gamma & -u^* \\
i\gamma_{12} & u^* & -u & -i\gamma_{21}
\end{array} \right)
\left(
\begin{array}{c}
\rho_{11} \\ \rho_{12} \\ \rho_{21} \\ \rho_{22}
\end{array} \right)
.
\end{eqnarray}
The interaction representation means that we have transformed the
mixed-state $\rho$ with the unitary transformation
$U=\textrm{diag}(1,e^{i\omega t},e^{-i\omega t},1)$. Since
$\textrm{Tr}[\rho]=1$, the density matrix $\rho$ can be
represented by the vector $q=^t(x,y,z)$ where $x=2\Re[\rho_{12}]$,
$y=2\Im[\rho_{12}]$ and $z=\rho_{22}-\rho_{11}$ and $q$ belongs to
the \emph{Bloch ball} $|q|\leq 1$. Equation (\ref{eqconc3}) takes
the form:
\begin{eqnarray}
\left\{ \begin{array}{lll}
\dot{x}=-\Gamma x+u_2z \\
\dot{y}=-\Gamma y-u_1z \\
\dot{z}=(\gamma_{12}-\gamma_{21})-(\gamma_{12}+\gamma_{21})z+u_1y-u_2x
\end{array} \right. .
\end{eqnarray}
$\Lambda=(\Gamma,\gamma_+,\gamma_-)$ is the set of parameters such
that $\gamma_+=\gamma_{12}+\gamma_{21}$ and
$\gamma_-=\gamma_{12}-\gamma_{21}$ and they satisfy the following
inequations $2\Gamma\geq \gamma_+\geq |\gamma_-|$ derived from the
Lindblad equation \cite{schirmer}, the Bloch ball $|q|\leq 1$
being invariant. The control is $u=u_1+iu_2$ where $u_1$ and $u_2$
are two real functions. We can write the control field
$u=|u|e^{i\phi}$ where $|u|\leq M$ and up to a rescaling of the
time and dissipative parameters we can assume that $|u|\leq 1$.

We consider the time-minimal transfer problem from a state $q_0$
to a state $q_1$. Hence, we have to analyze a time-minimal control
problem for a \emph{bilinear system} of the form:
\begin{equation}
\dot{q}=F_0(q)+\sum_{i=1}^2u_iF_i(q),~|u|\leq 1,\nonumber
\end{equation}
where the drift term $F_0$ depends upon three parameters. This
problem is a very difficult problem whose analysis requires
advanced mathematical tools from geometric control theory and
numerical simulations.

Such analysis is motivated by physical reasons. It is a
fundamental model in quantum control and more general problems can
be handled by coupling such systems. Numerous optimal control
results exist in the conservative case e.g. \cite{boscain2}, but
only partial ones for this problem: a pioneering work
\cite{sugnykontz} assuming $u$ real and a second one
\cite{bonnardsugny} for $u$ complex but restricted to
$\gamma_-=0$.

The objective of this article is double. First of all, we
introduce all the proper geometric tools to analyze the problem
using Pontryagin maximum principle. Secondly, our aim is to make a
complete study for every generic parameter in $\Lambda$, combining
mathematical reasonings and numerical simulations based on
\emph{shooting techniques} and including computations of
\emph{conjugate points} to test optimality. Based on the Cotcot
code \cite{BCT}, they can be used in practice to compute the true
optimal control, once the physical parameters are identified.

The organization of this article is the following. In section
\ref{sec2}, we complete the classification of the time-minimal
synthesis of \cite{sugnykontz} corresponding to the case where $u$
is real but introducing more general tools to handle the problem.
It corresponds to a time-minimal control problem of a
two-dimensional bilinear system in the single input case. The
optimal synthesis for a fixed initial point can be constructed by
gluing together local optimal syntheses. We can also make
estimates of switching points by lifting the system on a
semi-direct Lie group. \emph{This classification is physically
relevant to analyze the 3D-case because, using the symmetry of
revolution of the problem, it gives the time-minimal synthesis for
initial points $q_0=^t(0,0,\pm 1)$}. This geometric property is
explained in section \ref{sec3}. Moreover, using spherical
coordinates the system can be viewed as a system on a
\emph{two-sphere of revolution} coupled with the evolution of the
distance to the origin, which represents the \emph{purity} of the
system. According to the maximum principle, smooth extremals are
solutions of the Hamiltonian vector field $\vec{H}$ where
$H=H_0+(H_1^2+H_2^2)^{1/2}$, $H_i=\langle p,F_i(q)\rangle$, and
the control components are given by $u_i=H_i/(H_1^2+H_2^2)^{1/2}$,
$i=1,2$. Non smooth extremals can be constructed by connecting
smooth subarcs of the switching surface $\Sigma$: $H_1=H_2=0$. A
contribution of this article in section \ref{sec3} is to classify
the possible connections. We proved that every non smooth extremal
is either a solution of the 2D-single input system, assuming $u$
real, or occurs when meeting the equatorial plane of the Bloch
ball. In the second case, the switching can be handled numerically
using an integrator with an adaptative step. In the same section,
we combine analytical and numerical analysis to determine the
extremals and compute conjugate points. This completes the
analysis from \cite{bonnardsugny} in the integrable case. The
physical interpretation is presented as a conclusion.
\section{The 2D-case}\label{sec2}
Following \cite{sugnykontz}, a first step in the analysis is to
consider the following reduced system. Assuming $u$ real, the
$x$-coordinate is not controllable and we can consider the planar
single-input system:
\begin{eqnarray}
\dot{y}&=& -\Gamma y-u_1z\nonumber \\
\dot{z}&=& \gamma_--\gamma_+z+u_1y,|u_1|\leq 1\nonumber .
\end{eqnarray}
It gives the time-optimal analysis of the control problem when the
initial state $q(0)=(y(0),z(0))$ is a pure state on the $z$-axis,
that is $q(0)=(0,\pm 1)$. We proceed as follows to make the
analysis.
\subsection{Symmetry group}
A discrete symmetry group is associated to reflections with
respect to the different axes. More precisely, if $w=-z$ then one
 gets the same system changing $u_1$ into $-u_1$ and $\gamma_-$
 into $-\gamma_-$. If $w=-y$ then one gets the same system
 changing $u_1$ into $-u_1$. In particular, concerning the
 time-minimal control problem, this amounts to exchange the
 trajectories $\sigma_+$ and $\sigma_-$ corresponding respectively
 to $u_1=1$ and $u_1=-1$. Also, according to this property, the time-minimal synthesis is
symmetric with respect to the $z$-axis. This is connected to the
symmetry of revolution of the whole system which is explained
later.
\subsection{The feedback classification}
\begin{figure}
\centering
\includegraphics[width=0.4\textwidth]{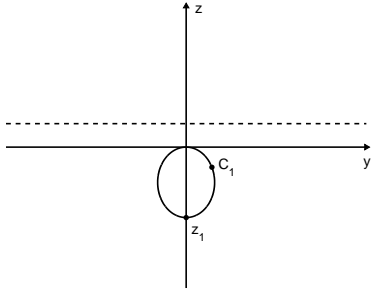}
\caption{\label{fig1} Diagram of the sets $S$ and $C$ in solid
lines for $\gamma_-=-0.2$, $\gamma_+=0.4$ and $\Gamma=1$. The
equation of the horizontal dashed line is
$z=\gamma_-/2(\gamma_+-\Gamma)$.}
\end{figure}

A preliminary step in our analysis is to consider the feedback
classification problem \cite{bonnardchyba}. The system is written
in a more compact form as follows:
\begin{equation}
\dot{q}=F(q)+uG(q)\nonumber
\end{equation}
where $F$ and $G$ are affine vector fields. To make the feedback
classification, we relax the control bound $|u|\leq 1$. The
geometric invariants are related to the sets:
\begin{itemize}
\item $S=\{q,\textrm{det}(G,[F,G])=0\}$ where are located the
singular trajectories.
\item $C=\{q,\textrm{det}(F,G)=0\}$ corresponding to the set of
points where $F$ and $G$ are collinear.
\end{itemize}
They are obtained by straightforward computations of Lie brackets:
\begin{eqnarray}
G=\left(\begin{array}{c}
-z \\
y
\end{array} \right),~
F=\left(\begin{array}{c}
-\Gamma y \\
\gamma_--\gamma_+z
\end{array} \right),~
[G,F]=\left(\begin{array}{c}
(\gamma_+-\Gamma)z-\gamma_- \\
(\gamma_+-\Gamma)y
\end{array} \right).\nonumber
\end{eqnarray}
Hence, $S$ is given by:
\begin{equation}
y[2(\Gamma-\gamma_+)z+\gamma_-]=0\nonumber
\end{equation}
and if $\gamma_+\neq \Gamma$, the singular set is defined by the
two lines $y=0$ and $z=\gamma_-/[2(\gamma_+-\Gamma)]$. The
collinear set is defined by:
\begin{equation}
\gamma_+z^2+\Gamma y^2-\gamma_-z=0\nonumber
\end{equation}
and is a closed curve formed by the union of two arcs of parabola,
containing $(0,0)$ and $(z_1,0)$, with $z_1=\gamma_-/\gamma_+$,
which is the equilibrium state of the free motion. More generally,
$C$ contains the equilibrium points for the dynamics with constant
control $u_0$, since $F+u_0G=0$. In particular, the equilibrium
point when $u_1=1$ is given by
\begin{equation}
C_1:~y=\frac{\gamma_-}{1+\gamma_+\Gamma},~z=-\Gamma y.\nonumber
\end{equation}
The collinear set $C$ shrinks into a point when $\gamma_-=0$.
Computing the intersection of $C$ with the singular line
$z=\gamma_-/[2(\gamma_+-\Gamma)]$, one gets:
\begin{equation}
\Gamma
y^2=\frac{\gamma_-^2}{4(\gamma_+-\Gamma)^2}(\gamma_+-2\Gamma).\nonumber
\end{equation}
If $\gamma_-\neq 0$, since $2\Gamma\geq \gamma_+$, one deduces
that the intersection is empty except in the case where
$\gamma_+=2\Gamma$. An important consequence is to simplify the
classification of the optimal syntheses. We represent on Fig.
\ref{fig1} the sets $S$ and $C$ for a situation with $\gamma_-<0$
and $\gamma_+-\Gamma<0$.
Another feedback invariant is the optimality status of singular
trajectories. Using the generalized Legendre-Clebsch condition, it
is splitted into fast and slow directions. In the 2D-case, it is
tested by Lie brackets configurations and can be computed by
introducing:
\begin{equation}
D=\textrm{det}(G,[[G,F],G]),~D''=\textrm{det}(G,F).\nonumber
\end{equation}
The trajectory is time-optimal in the so-called hyperbolic case
$DD''>0$ and time-maximal in the so-called elliptic case $DD''<0$.
Computing Lie brackets of length 3, we have:
\begin{eqnarray}
[[G,F],G]=\left(\begin{array}{c}
2(\gamma_+-\Gamma)y \\
\gamma_--2(\gamma_+-\Gamma)z
\end{array} \right)\nonumber
\end{eqnarray}
and
\begin{eqnarray}
[[G,F],F]=\left(\begin{array}{c}
(\gamma_+-\Gamma)(\gamma_--\gamma_+z)+\Gamma[(\gamma_+-\Gamma)z-\gamma_-] \\
(\gamma_+-\Gamma)^2y
\end{array} \right).\nonumber
\end{eqnarray}
Hence:
\begin{eqnarray}
D''=\left|\begin{array}{cc}
-z & -\Gamma y \\
y & \gamma_--\gamma_+z
\end{array} \right|,~ D=\left|\begin{array}{cc}
-z & 2(\gamma_+-\Gamma)y \\
y & \gamma_-2(\gamma_+-\Gamma)z
\end{array} \right|.\nonumber
\end{eqnarray}
For the singular direction $y=0$, we get:
\begin{equation}
DD''=2z^2(\gamma_+-\Gamma)\gamma_+(z-\frac{\gamma_-}{2(\gamma_+-\Gamma)})(z-\frac{\gamma_-}{\gamma_+}).\nonumber
\end{equation}
Near the origin, the sign is always positive if $\gamma_-\neq 0$.
If $\gamma_-=0$, the sign is given by $(\gamma_+-\Gamma)$. For the
singular direction $z=\gamma_-/[2(\gamma_+-\Gamma)]$, we have:
\begin{equation}
DD''=\frac{y^2}{2(\gamma_+-\Gamma)}[\gamma_-^2(\gamma_+-2\Gamma)-4\Gamma
y^2(\gamma_+-\Gamma)^2].\nonumber
\end{equation}
Hence, near the origin, the optimality is given by the sign of
$(\gamma_+-2\Gamma)(\gamma_+-\Gamma)$. Moreover, since $\Gamma\geq
\gamma_+/2$, one gets that $DD''>0$ if $\gamma_+-\Gamma<0$ and
$DD''<0$ if $\gamma_+-\Gamma>0$.
\subsection{Computation of a
normal form} In order to analyze the time-optimal control, we
compute a global normal form up to a polar singularity for the
action of the feedback group. A first step is to linearize the
vector field $G$ since it is connected to the evaluation of the
switching times. One further normalization is to straight the
horizontal singular line: $z=\gamma_-/[2(\gamma_+-\Gamma)]$. Using
polar coordinates:
\begin{equation}
y=\rho\cos\phi ,~ z=\rho\sin\phi ,\nonumber
\end{equation}
one gets:
\begin{eqnarray}
\dot{\rho}&=& \gamma_-\sin\phi+\rho[-\Gamma+(\Gamma-\gamma_+)\sin^2\phi]\nonumber\\
\dot{\phi}&=&
\frac{\gamma_-\cos\phi}{\rho}+(\Gamma-\gamma_+)\frac{\sin(2\phi)}{2}+u_1\nonumber.
\end{eqnarray}
If we use the coordinates $x=\rho^2/2$ and $z$, the system
becomes:
\begin{eqnarray}
\dot{x}&=& -2\Gamma x+\gamma_-z+z^2(\Gamma-\gamma_+)\nonumber\\
\dot{z}&=& \gamma_--\gamma_+z+u_1\sqrt{2x-z^2}\nonumber.
\end{eqnarray}
Making a feedback transformation of the form $u_1\to \beta u_1$
where $\beta$ is a function of $x$ and $z$, we can consider the
system:
\begin{eqnarray}
\dot{x}&=& -2\Gamma x+\gamma_-z+z^2(\Gamma-\gamma_+)\nonumber\\
\dot{z}&=& \gamma_--\gamma_+z+u_1\nonumber.
\end{eqnarray}
If we set $z=Z+z_0$ where $z_0=\gamma_-/[2(\gamma_+-\Gamma)]$, we
obtain the system:
\begin{eqnarray}
\dot{x}&=& \frac{\gamma_-^2}{4(\gamma_+-\Gamma)}-2\Gamma x+(\Gamma-\gamma_+)z^2\nonumber\\
\dot{z}&=&
\frac{\gamma_-(\gamma_+-2\Gamma)}{2(\gamma_+-\Gamma)}-\gamma_+z+u_1
\nonumber.
\end{eqnarray}
In this simplified model where the control is rescaled by the
positive parameter $\beta$, we keep most of the information about
the initial system. In particular, all the feedback invariants are
preserved: the collinear set corresponds to $\dot{x}=0$ and the
singular set is identified to $z=0$ with its optimality status,
while we have wiped out the singular line $y=0$. Due to the
feedback transformation, we lose however the singularities of the
vector fields $F+G$ and $F-G$ and also the saturation phenomenon
of the singular control.

For the simplified model, the adjoint system takes the form:
\begin{eqnarray}
\dot{p}_x&=& 2\Gamma p_x\nonumber\\
\dot{p}_z&=& -2zp_x(\Gamma-\gamma_+)+p_z\gamma_- \nonumber,
\end{eqnarray}
and can be easily integrated to compute the time-minimal
synthesis.
\subsection{The saturation phenomenon}
One interesting property which is not captured by the normal form
is when the singular control is saturating along the horizontal
singular line $z=\gamma_-/[2(\gamma_+-\Gamma)]$. Introducing
$D'=\textrm{det}(G,[[G,F],F])$, we get on the singular line:
$D'=y\gamma_-(2\Gamma-\gamma_+)$. The singular control is given
by:
\begin{equation}
u_s=-\frac{D'}{D}=\frac{\gamma_-(\gamma_+-2\Gamma)}{2y(\Gamma-\gamma_+)}
,\nonumber
\end{equation}
and saturation occurs when $|u_s|=1$. Observe that if
$\gamma_-(2\Gamma-\gamma_+)\neq 0$ then the singular control is
never admissible when $y=0$.
\subsection{The switching function}
For the true system, the switching function is more intricate but
can be still analyzed using geometric and numerical techniques.
The switching function $\Phi$ is given by $\Phi(t)=p(t)G(q(t))$
and switching occurs when $\Phi(t)=0$. By construction, $G$ is
tangent to the circle $S^1$, hence is rotating when we follow an
arc curve $\sigma_+$ or $\sigma_-$. The dynamics of $p$ is given
by the adjoint equation:
\begin{equation}
\dot{p}=-p(\frac{\partial F}{\partial q}+u\frac{\partial
G}{\partial q}),~u=\pm 1 .\nonumber
\end{equation}
The corresponding dynamics is linear and $p$ can be either
oscillating if the eigenvalues are complex or non oscillating if
they are real.

An equivalent but more geometric test is the use of the standard
$\theta$-function introduced in \cite{boscainpiccoli} and defined
as follows. Let $v$ be the tangent vector solution of the
variational equation:
\begin{equation}
\dot{v}=(\frac{\partial F}{\partial q}+u\frac{\partial G}{\partial
q})v,~u=\pm 1 ,\nonumber
\end{equation}
whose dynamics is similar to the one of the adjoint vector. Let
$t_1<t_2$ be two consecutive switching times on an arc $\sigma_+$
or $\sigma_-$. One can set $t_1=0$ and $t_2=t$. By definition, we
have:
\begin{equation}
p(0)G(q(0))=p(t)G(q(t))=0 .\nonumber
\end{equation}
We denote by $v(\cdot)$ the solution of the variational equation
such that $v(t)=G(q(t))$ and where this equation is integrated
backwards from time $t$ to time 0. By construction $p(0)v(0)=0$
and we deduce that at time 0, $p(0)$ is orthogonal to $G(q(0))$
and to $v(0)$. Therefore, $v(0)$ and $G(q(0))$ are collinear;
$\theta(t)$ is defined as the angle between $G(q(0))$ and $v(0)$
measured counterclockwise. One deduces that switching occurs when
$\theta(t)=0~[\pi]$. In the analytic case, $\theta(t)$ can be
computed using Lie brackets. Indeed, for $u=\varepsilon$,
$\varepsilon=\pm 1$, we have by definition
\begin{equation}
v(0)=e^{-t\textrm{ad}(F+\varepsilon G)}G(q(t)),\nonumber
\end{equation}
and in the analytic case, the ad-formulae gives :
\begin{equation}
v(0)=\sum_{n\geq 0}\frac{(-t)^n}{n!}\textrm{ad}^n(F+\varepsilon
G)G(q(t)).\nonumber
\end{equation}
Here, to make the computation explicit, we take advantage of the
fact that we can lift our bilinear system into an invariant system
onto the semi-direct Lie group $GL(2,\mathbb{R})\times_S
\mathbb{R}^2$ identified to the set of matrices of
$GL(3,\mathbb{R})$:
\begin{eqnarray}
\left(\begin{array}{cc}
1 & 0 \\
g & v
\end{array} \right),~ g\in GL(2,\mathbb{R}),~ v\in\mathbb{R}^2,\nonumber
\end{eqnarray}
acting on the subspace of vectors in $\mathbb{R}^3$:
$\left(\begin{array}{c}
1 \\
q
\end{array} \right)$.

Lie brackets computations are defined as follows. We set:
\begin{equation}
F(q)=Aq+a,~G(q)=Bq,\nonumber
\end{equation}
and $F,G$ are identified to $(A,a)$, $(B,0)$ in the Lie algebra
$\underline{gl}(2,\mathbb{R})\times \mathbb{R}^2$. The Lie
brackets computations on the semi-direct product are defined by:
\begin{equation}
[(A',a'),(B',b')]=([A',B'],A'b'-B'a').\nonumber
\end{equation}
We now compute $\exp[-t\textrm{ad}(F+\varepsilon G)]$. The first
step consists in determining $\exp[-t\textrm{ad}(A+\varepsilon
B)]$ which amounts to compute $\textrm{ad}(A+\varepsilon B)$. We
write $\underline{gl}(2,\mathbb{R})=c\oplus
\underline{sl}(2,\mathbb{R})$ where $c$ is the center
\begin{eqnarray}
\mathbb{R}\left(\begin{array}{cc}
1 & 0 \\
0 & 1
\end{array} \right).\nonumber
\end{eqnarray}
We choose the following basis of $\underline{sl}(2,\mathbb{R})$:
\begin{eqnarray}
B=\left(\begin{array}{cc}
0 & -1 \\
1 & 0
\end{array} \right),~C=\left(\begin{array}{cc}
0 & 1 \\
1 & 0
\end{array} \right)~\textrm{and}~D=\left(\begin{array}{cc}
1 & 0 \\
0 & -1
\end{array} \right).\nonumber
\end{eqnarray}
The matrix $A$ is decomposed into:
\begin{eqnarray}
A=\left(\begin{array}{cc}
-\Gamma & 0 \\
0 & -\gamma_+
\end{array} \right)=\left(\begin{array}{cc}
\lambda & 0 \\
0 & \lambda
\end{array} \right)+\left(\begin{array}{cc}
s & 0 \\
0 & -s
\end{array} \right)\nonumber
\end{eqnarray}
and hence $\lambda=-(\Gamma+\gamma_+)/2$ and
$s=(\gamma_+-\Gamma)/2$. In the basis ($B$, $C$, $D$),
$\textrm{ad}(A+\varepsilon B)$ is represented by the matrix:
\begin{eqnarray}
\left(\begin{array}{ccc}
0 & -2s & 0 \\
-2s & 0 & 2\varepsilon\\
0 & -2\varepsilon & 0
\end{array} \right).\nonumber
\end{eqnarray}
The characteristic polynomial is
$P(\lambda)=-\lambda(\lambda^2+4(\varepsilon^2-s^2))$ and the
eigenvalues are $\lambda=0$ and $\lambda_i=\pm
2\sqrt{s^2-\varepsilon^2}$, $i=1,2$; $\lambda_1$ and $\lambda_2$
are distinct and real if $|\gamma_+-\Gamma|>2$ and we note
$\lambda_1=2\sqrt{\varepsilon^2-s^2}$, $\lambda_2=-\lambda_1$;
$\lambda_1$ and $\lambda_2$ are distinct and imaginary if
$|\gamma_+-\Gamma|<2$ and we note
$\lambda_1=2i\sqrt{\varepsilon^2-s^2}$, $\lambda_2=-\lambda_1$. To
compute $e^{-t\textrm{ad}(A+\varepsilon B)}$, we must distinct two
cases.\\
\textbf{Real case:} In the basis $B$, $C$, $D$, the eigenvectors
corresponding to $\{0,\lambda_1,\lambda_2\}$ are respectively:
$v_0= ^t(\varepsilon,0,s)$, $v_1= ^t(2s,-\lambda_1,2\varepsilon)$
and $v_2= ^t(2s,-\lambda_2,2\varepsilon)$. Therefore, in this
eigenvector basis, $\exp[-t\textrm{ad}(A+\varepsilon B)]$ is the
diagonal matrix: $\textrm{diag}(1,e^{-\lambda_1
t},e^{-\lambda_2t})$. To compute $\exp[-t\textrm{ad}(A+\varepsilon
B)]B$, we use the decomposition
\begin{equation}
B=\alpha v_0+\beta v_1+\gamma v_2,\nonumber
\end{equation}
with:
\begin{equation}
\alpha=\frac{\varepsilon}{\varepsilon^2-s^2},~\beta=\frac{-\lambda_2s}{2(\lambda_2-\lambda_1)(\varepsilon^2-s^2)},~
\gamma=\frac{-\lambda_1s}{2(\lambda_1-\lambda_2)(\varepsilon^2-s^2)}.\nonumber
\end{equation}
Hence one gets:
\begin{equation}
e^{-t\textrm{ad}(A+\varepsilon B)}B=\alpha v_0+\beta e^{-\lambda_1
t}v_1+\gamma e^{-\lambda_2 t}v_2.\nonumber
\end{equation}
To test the collinearity at $q_0$, we compute
\begin{equation}
\textrm{det}(B(q_0),e^{-t\textrm{ad}(A+\varepsilon
B)}B(q_0))=0\nonumber
\end{equation}
where the determinant is equal to
\begin{equation}\label{eq22}
(z_0^2-y_0^2)(\alpha s+2\varepsilon(\beta e^{-\lambda_1 t}+\gamma
e^{-\lambda_2 t}))+2y_0z_0(\lambda_1\beta e^{-\lambda_1
t}+\lambda_2\gamma e^{-\lambda_2 t}).\nonumber
\end{equation}
Using the fact that this last expression has at most two zeros,
one being for $t=0$, it is straightforward to check that for
$q_0=^t(0,1)$, this expression only vanishes at $t=0$. This proves
the result numerically checked in \cite{sugnykontz} that there is
\emph{at most one switching}. This analysis can be generalized to
any initial condition.
\\
\textbf{Imaginary case:} In this case, we note $\lambda_1=i\theta$
 the eigenvalue associated to the eigenvector
$^t(2s,-i\theta,2\varepsilon)$. We consider the real part
$v_1=^t(2s,0,2\varepsilon)$ and the imaginary part
$v_2=^t(0,-\theta,0)$. In the basis $v_0=^t(\varepsilon,0,s)$,
$v_1$, $v_2$, $\textrm{ad}(A+\varepsilon B)$ takes the normal
form:
\begin{eqnarray}
\textrm{diag}(0,\left(\begin{array}{cc}
0 & \theta \\
-\theta & 0
\end{array} \right) ).\nonumber
\end{eqnarray}
Hence, we have in this basis:
\begin{eqnarray}
e^{-t\textrm{ad}(A+\varepsilon
B)}=\textrm{diag}(1,\left(\begin{array}{cc}
\cos(\theta t) & -\sin(\theta t) \\
\sin(\theta t) & \cos(\theta t)
\end{array} \right) ).\nonumber
\end{eqnarray}
We decompose $B$ in the same basis:
\begin{equation}
B=\alpha v_0+\beta v_1+\gamma v_2,\nonumber
\end{equation}
where
\begin{equation}
\alpha=\frac{\varepsilon}{\varepsilon^2-s^2},~\beta=-\frac{s}{2(\varepsilon^2-s^2)},~\gamma=0.\nonumber
\end{equation}
Hence, we get:
\begin{equation}
e^{-t\textrm{ad}(A+\varepsilon B)}B=\alpha v_0+\beta [\cos(\theta
t)v_1+\sin(\theta t)v_2].\nonumber
\end{equation}
Computing we obtain:
\begin{equation}
\textrm{det}(Bq_0,e^{-t\textrm{ad}(A+\varepsilon
B)}B(q_0))=(z_0^2-y_0^2)(\alpha s+2\varepsilon \beta\cos(\theta
t))+2\beta\theta\sin(\theta t)y_0 z_0\nonumber
\end{equation}
which vanishes for two values of $t$ in $[0,2\pi/\theta[$ if
$y_0\neq z_0$. These two values coincide for $q_0=^t(0,1)$. Hence
\emph{the switchings occur periodically with a period of
$2\pi/\theta$} which confirms the numerical simulations of
\cite{sugnykontz}.\\
\textbf{Case $\gamma_-\neq 0$:} The Lie bracket is given by:
\begin{equation}
[(A',a'),(B',b')]=([A',B'],A'b'-B'a').\nonumber
\end{equation}
We note $(e_1,e_2)$ the $\mathbb{R}^2$-canonical basis and
$a=\gamma_- e_2$. We have:
\begin{eqnarray}
\textrm{ad}(A+\varepsilon B,a)\cdot (B,0)&=& ([A+\varepsilon
B,B],-Ba)\nonumber\\
&=& ([A,B],-Ba)\nonumber ,
\end{eqnarray}
\begin{eqnarray}
\textrm{ad}^2(A+\varepsilon B,a)\cdot (B,0)&=& [(A+\varepsilon
B,a),([A,B],-Ba)]\nonumber\\
&=& (\textrm{ad}^2(A+\varepsilon B,B),-(A+\varepsilon
B)Ba-[A,B]a)\nonumber .
\end{eqnarray}
More generally, one gets:
\begin{equation}
\textrm{ad}^k(A+\varepsilon B,a)\cdot (B,0)=(\textrm{ad}^k
(A+\varepsilon B)\cdot B,v_k)\nonumber
\end{equation}
where $v_k$ is given by the recurrence relation:
\begin{equation}
v_k=-\textrm{ad}^{k-1}(A+\varepsilon B)\cdot Ba+(A+\varepsilon
B)v_{k-1}.\nonumber
\end{equation}

The computation is intricate but simplifies if $\gamma_+=\Gamma$
since in this case $\textrm{ad}^k(A+\varepsilon B)\cdot B=0$ for
$k\geq 1$. Numerical simulations have to be used to compute the switchings sequence.\\
\textbf{Generalization}: This technique can be generalized to the
time-minimal control problem in the full control case, replacing
the control domain $|u|\leq 1$ by $|u_1|,|u_2|\leq 1$.
\subsection{The time-minimal
synthesis} 
\begin{figure}
\centering
\includegraphics[width=0.5\textwidth]{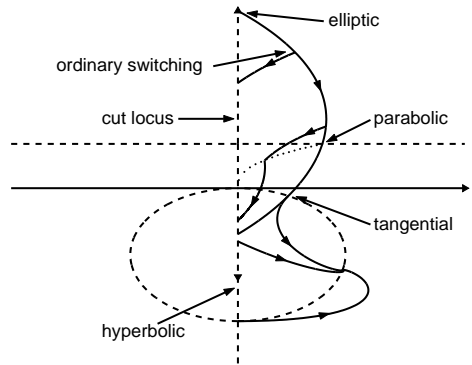}
\caption{\label{fig3} Cut and switching loci for the case
$\gamma_-<0$. The sets $C$ and $S$ are represented in dashed
lines, the switching locus in dotted lines.}
\end{figure}

\begin{figure}
\centering
\includegraphics[width=0.4\textwidth]{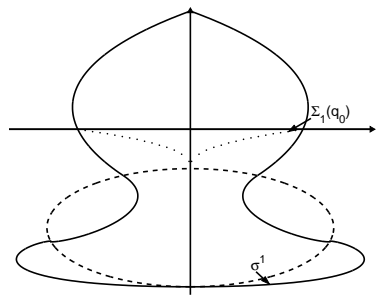}
\caption{\label{fig4}Poisson shape of the accessibility set.}
\end{figure}

We use \cite{bonnardchyba} as general reference on
time-minimal synthesis, see also \cite{boscainpiccoli}. The
initial condition is fixed to $q_0=^t(0,1)$ and we consider the
problem of constructing the time-minimal synthesis from this
initial point. This amounts to compute two objects:
\begin{itemize}
\item The switching locus $\Sigma (q_0)$ of optimal trajectories
which is deduced from the switching locus of extremal
trajectories.
\item The cut locus $C(q_0)$ which is formed by the set of points
where a minimizer ceases to be optimal.
\end{itemize}
In order to achieve this task, we must glue together local time
minimal syntheses which are classified. To be more precise, we
recover the case (d) of \cite{sugnykontz}, the gluing being
indicated on Fig. \ref{fig3} on which we have represented the
local extremal classifications of \cite{bonnardchyba} which are
crucial to deduce the optimal syntheses.
In this case, the cut locus is a segment of the $z-$axis and its
birth is located at the initial point $(0,1)$ which is a
consequence of the elliptic situation. The switching locus is the
union of a fast singular trajectory, corresponding to an
hyperbolic point and a curve $\Sigma_1(q_0)$ corresponding to
parabolic points (for the terminology see \cite{bonnardchyba}).
Note also the importance of the tangential point where arcs
$\sigma_+$ and $\sigma_-$ are tangent leading to the fish-shaped
accessibility set $A^+(q_0)$ represented on Fig. \ref{fig4}. This
set is not closed since the arc $\sigma^1$ starting from $(0,z_1)$
is not in $A^+(q_0)$.
We next list the micro-local situations we need to construct the
synthesis.
\begin{itemize}
\item \textbf{Ordinary switching points}: The local synthesis is
given by $\sigma_-\sigma_+$ or $\sigma_+\sigma_-$. The two cases
are distinguished using for instance the clock form $\omega=pdq$
with $\langle p,G\rangle =0$ and $\langle p,F\rangle =1$ which is
also useful to get more global results.
\item \textbf{Hyperbolic case}: Existence of a fast singular
admissible trajectory. The optimal synthesis is of the form
$\sigma_\pm\sigma_s\sigma_\pm$ where $\sigma_s$ is a singular arc
(see Fig. \ref{fig5a}).
\begin{figure}
\centering
\includegraphics[width=0.4\textwidth]{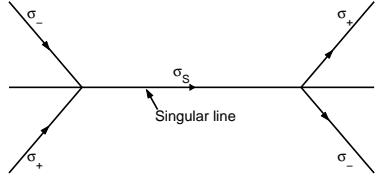}
\caption{\label{fig5a} Structure of the extremals for the
hyperbolic case.}
\end{figure}
\item \textbf{Elliptic case}: Existence of a slow admissible singular
trajectory. An optimal arc is bang-bang with at most one
switching. Not every extremal trajectory is optimal and we have
birth of a cut locus (see Fig. \ref{fig5}).
\begin{figure}
\centering
\includegraphics[width=0.4\textwidth]{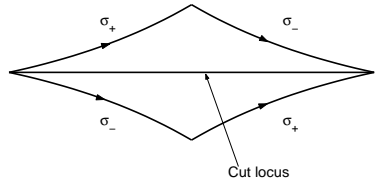}
\caption{\label{fig5} Elliptic case.}
\end{figure}
\item \textbf{Parabolic point}: It corresponds to a non-admissible
singular direction. Every extremal curve is bang-bang with at most
two switchings. In our case, the initial point is fixed and the
switching locus starts with the intersection of $\sigma_-$ with
the singular line (see Fig. \ref{fig6}).
\begin{figure}
\centering
\includegraphics[width=0.4\textwidth]{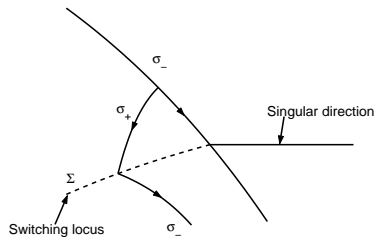}
\caption{\label{fig6} Parabolic case.}
\end{figure}
\item \textbf{Saturating case}:\\
A \emph{fast} singular trajectory is saturating at a point $M$:
birth of a switching curve at $M$ (see Fig. \ref{fig6a}).
\begin{figure}
\centering
\includegraphics[width=0.4\textwidth]{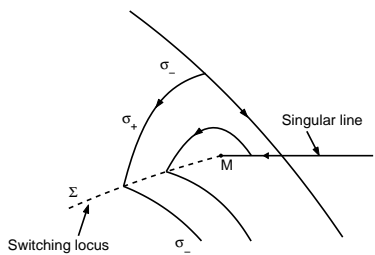}
\caption{\label{fig6a} Saturating case}
\end{figure}
\item \textbf{A $C\cap S\neq \emptyset$ case}:\\
A \emph{fast} singular trajectory meets the set $C$ and becomes
\emph{slow}.
\begin{figure}
\centering
\includegraphics[width=0.4\textwidth]{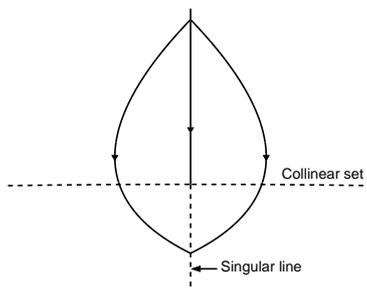}
\caption{\label{fig7a} A $C\cap S\neq \emptyset$ case}
\end{figure}
\end{itemize}
\subsection{Classification of the optimal syntheses}
We describe the different time-optimal syntheses in the
single-input case. Without loss of generality, we restrict the
study to the initial point $q_0=^t(0,1)$. The classification is
done with respect to the relative positions of the feedback
invariants $C$ and $S$ and to the optimal status of singular
extremals which are fast or slow according to the values of
$\Gamma$, $\gamma_+$ and $\gamma_-$.

\begin{figure}
\centering
\includegraphics[width=0.4\textwidth]{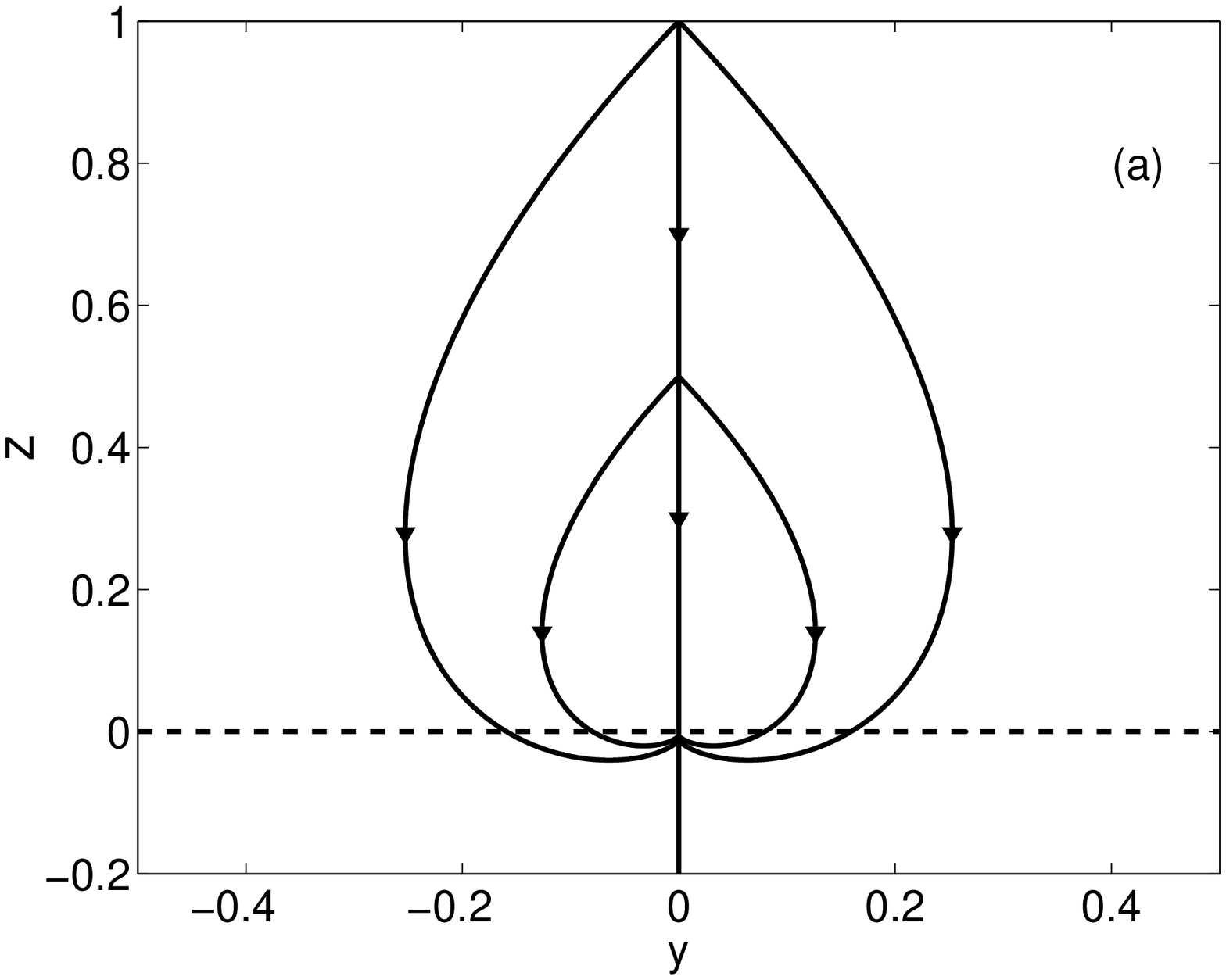}
\includegraphics[width=0.4\textwidth]{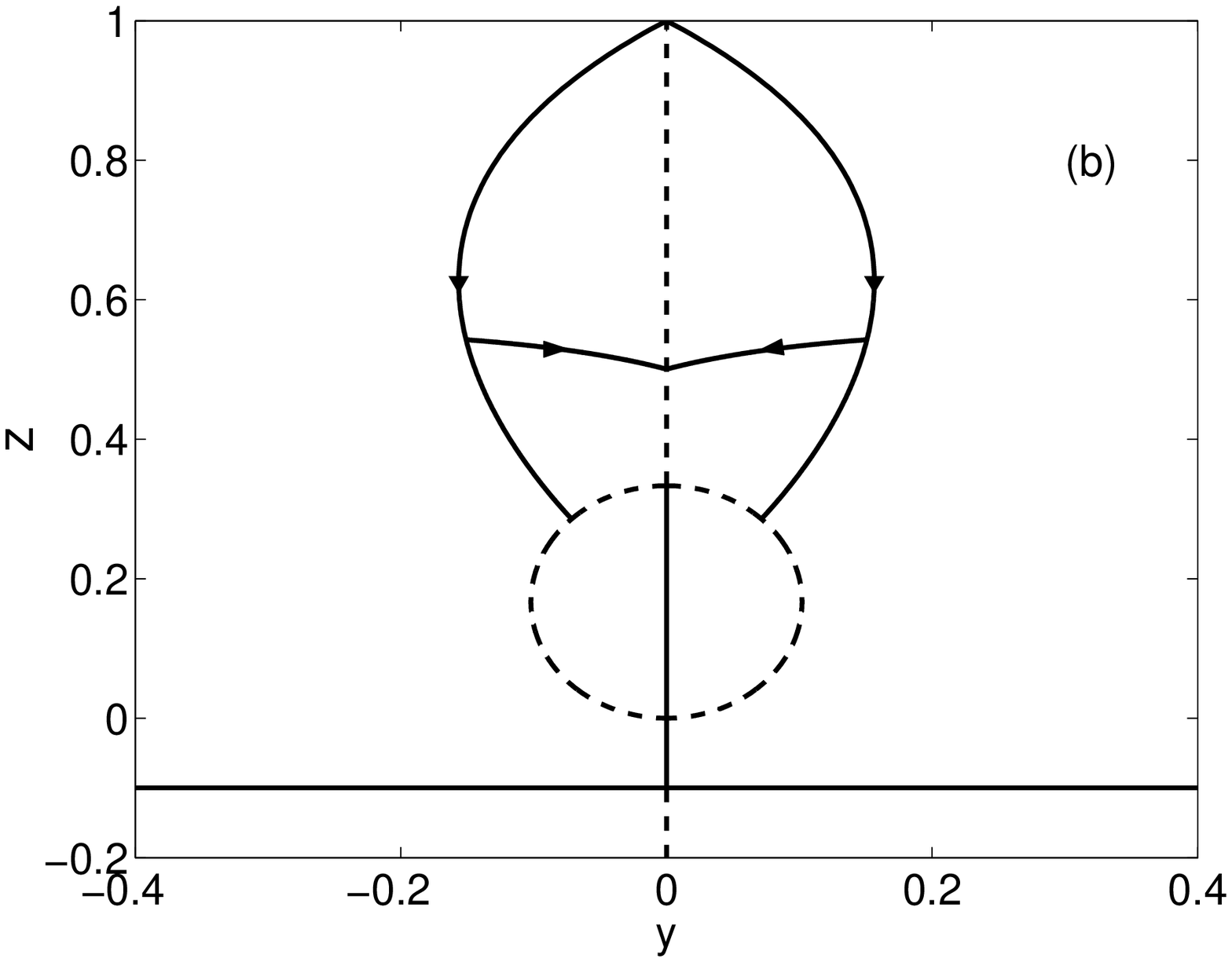}
\includegraphics[width=0.4\textwidth]{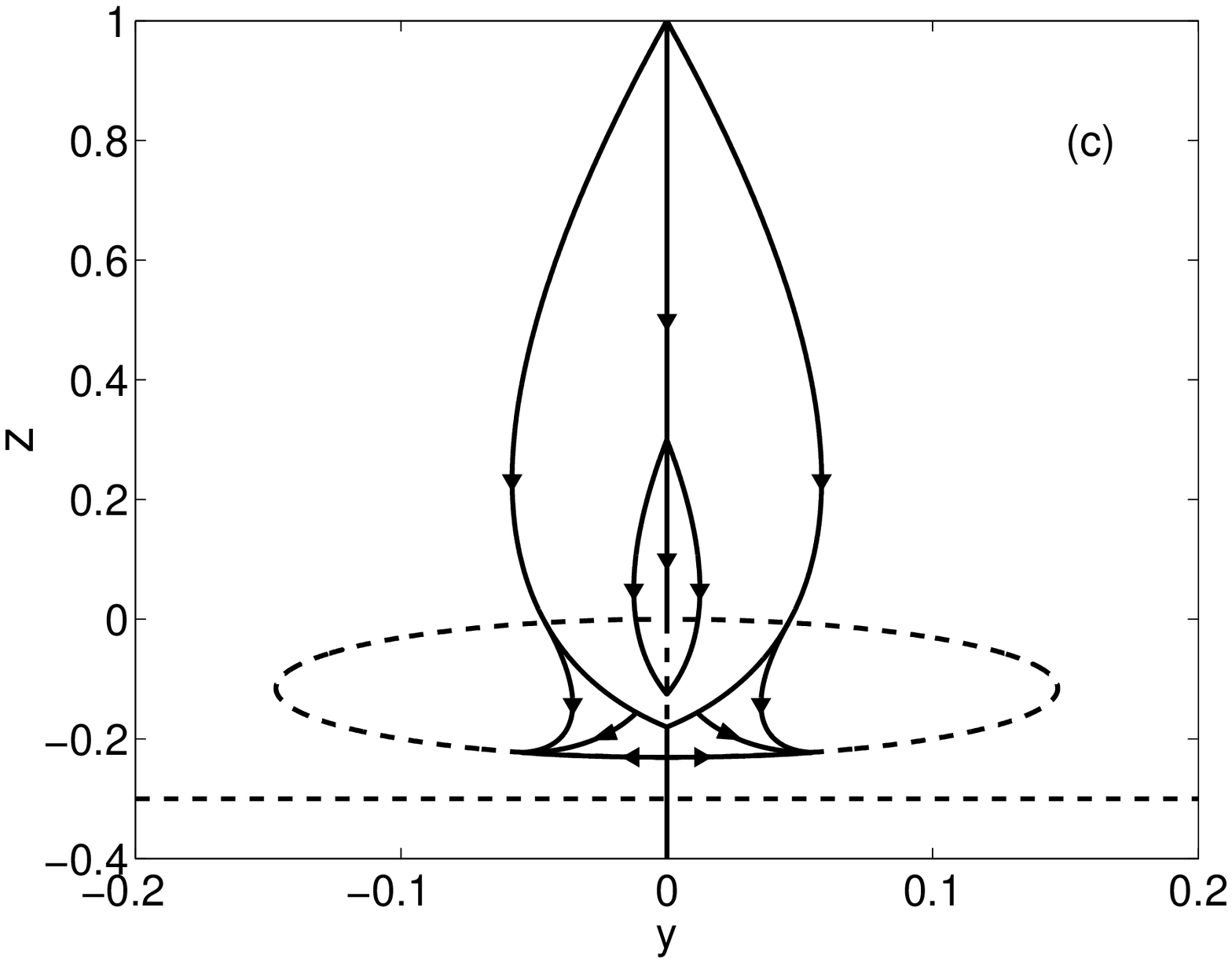}
\includegraphics[width=0.4\textwidth]{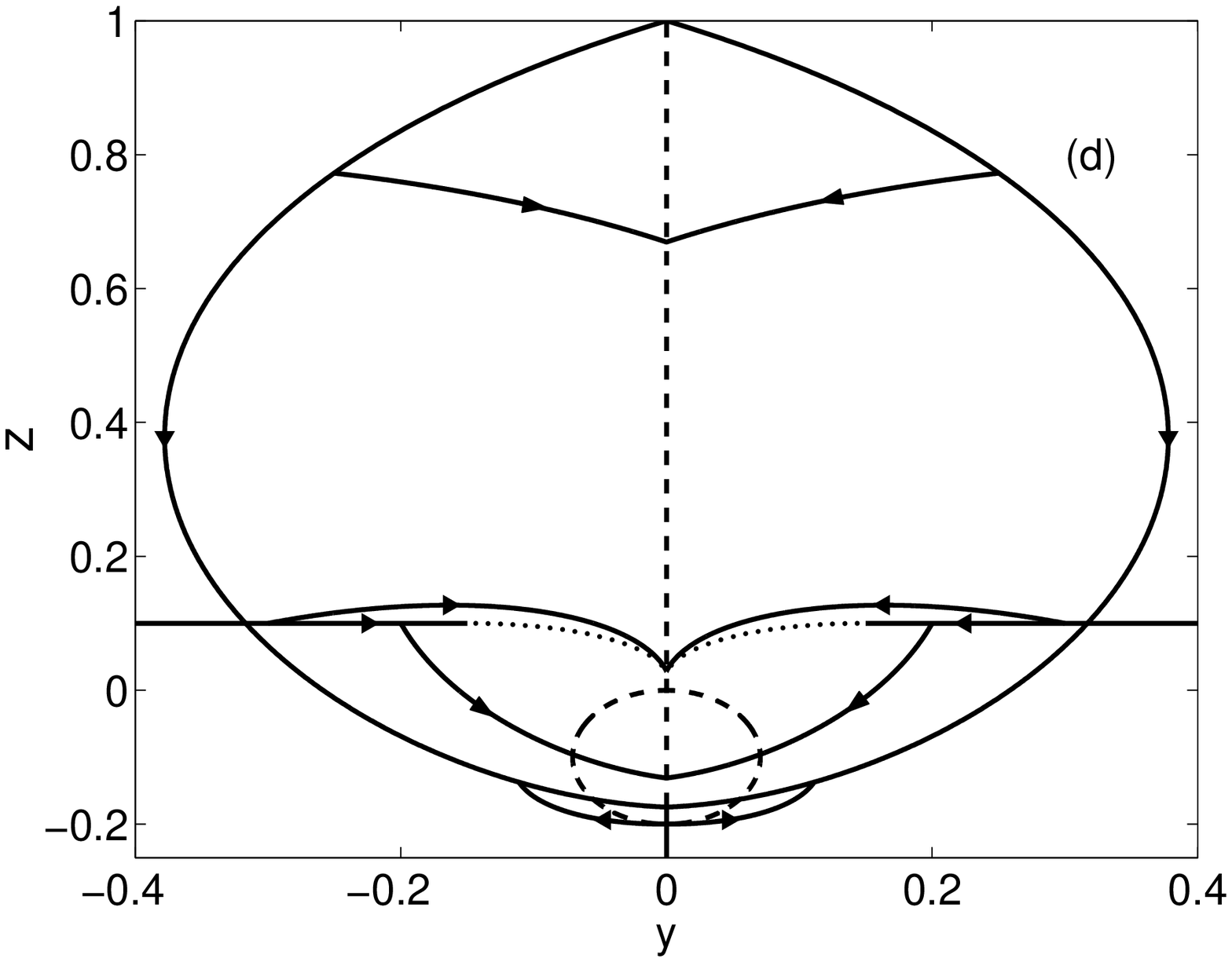}
\caption{\label{figsynt} Optimal syntheses for (a) ($\Gamma=1.1$,
$\gamma_+=1.6$, $\gamma_-=0$), (b) ($\Gamma=4$, $\gamma_+=1.5$,
$\gamma_-=0.5$), (c) ($\Gamma=4$, $\gamma_+=6.5$, $\gamma_-=-1.5$)
and (d) ($\Gamma=1$, $\gamma_+=0.5$, $\gamma_-=-0.1$). Solid and
dashed vertical and horizontal lines correspond respectively to
fast and slow singular lines. The set $C$ is represented in dashed
lines. The switching locus is plotted in dotted line. In (d), only
the admissible singular horizontal line is represented in solid
line.}
\end{figure}

For $\gamma_-=0$, the set $C$ is restricted to the origin and we
have two cases according to the sign of $\Gamma-\gamma_+$. Note
that the form of the extremals $\sigma_+$ and $\sigma_-$ starting
from $q_0$ depends on the sign of $|\Gamma-\gamma_+|-2$. Two cases
for $\Gamma>\gamma_+$ are presented in \cite{sugnykontz}. We
complete this study with the optimal synthesis for
$\Gamma<\gamma_+$ and $|\Gamma-\gamma_+|<2$ displayed in Fig.
\ref{figsynt}a.

For $\gamma_-\neq 0$, we distinguish four cases according to the
signs of $\gamma_-$ and $\Gamma-\gamma_+$. One case
($\Gamma>\gamma_+$ and $\gamma_-<0$) is treated in
\cite{sugnykontz}. We consider here three types of optimal
synthesis represented in Fig. \ref{figsynt}b, \ref{figsynt}c and
\ref{figsynt}d. Note that in a same class of synthesis the
reachable set from the initial point $q_0$ depends on the
dissipative parameters which can modify the structure of the
synthesis. The last case $\gamma_->0$ and $\gamma_+>\Gamma$ can be
deduced from the case $\gamma_->0$ and $\gamma_+<\Gamma$ since the
horizontal singular line plays no role in both cases. The
synthesis of Fig. \ref{figsynt}d is very similar to the one of
Fig. \ref{fig3} except the fact that a part of the horizontal
singular line is admissible. The switching locus has been computed
numerically using the switching function $\Phi$.

The role of the parameter $\gamma_-$ is clearly illustrated in
Figs. \ref{figsynt}a and \ref{figsynt}c. The case $\gamma_-=0$ is
a degenerate case where the set $C$ shrinks into a point. The
variation of $\gamma_-$ induces a \emph{bifurcation} of the
control system leading to new structures of the optimal synthesis.
For $\gamma_-\neq 0$, the set $C$ is the union of two branches of
parabola. The optimal status of the vertical singular line changes
when this line crosses the set $C$ in Fig. \ref{figsynt}c.
\section{The bi-input case}\label{sec3}
\subsection{Geometric analysis}
The system is written in short in Cartesian coordinates as
follows:
\begin{equation}
\dot{q}=F_0(q)+u_1F_1(q)+u_2F_2(q),~|u|\leq 1.\nonumber
\end{equation}
Introducing the Hamiltonians $H_i=\langle p,F_i\rangle,~i=0,1,2$,
the pseudo-Hamiltonian associated to the time-optimal control
problem is:
\begin{equation}
H=H_0+\sum_{i=1}^2u_iH_i+p_0,\nonumber
\end{equation}
where $p_0\leq 0$. The time-optimal control is given outside the
switching surface $\Sigma$: $H_1=H_2=0$, by
$u_i=H_i/\sqrt{H_1^2+H_2^2}$, $i=1,2$, with the corresponding true
Hamiltonian:
\begin{equation}
H_r=H_0+\sqrt{H_1^2+H_2^2},\nonumber
\end{equation}
whose solutions (outside $\Sigma$) are smooth and are called
\emph{extremals of order 0}. More general non smooth extremals can
be obtained by connecting such arcs through $\Sigma$.

To make the geometric analysis and to highlight the symmetry of
revolution, the system is written using the spherical coordinates:
\begin{equation}
x=\rho\sin\phi\cos\theta,~y=\rho\sin\phi\sin\theta,~z=\rho\cos\phi\nonumber
\end{equation}
and a feedback transformation:
\begin{equation}
v_1=u_1\cos\theta+u_2\sin\theta,~v_2=-u_1\sin\theta+u_2\cos\theta
.\nonumber
\end{equation}
We obtain the system:
\begin{subequations}\label{eq31}
\begin{eqnarray}
\dot{\rho}&=&\gamma_-\cos\phi-\rho(\gamma_+\cos^2\phi+\Gamma\sin^2\phi) \label{eq31a}\\
\dot{\phi}&=&-\frac{\gamma_-\sin\phi}{\rho}+\frac{\sin(2\phi)}{2}(\gamma_+-\Gamma)+v_2 \label{eq31b}\\
\dot{\theta} &=& -\cot\phi v_1 \label{eq31c}.
\end{eqnarray}
\end{subequations}
Hence, one deduces that the true Hamiltonian is:
\begin{equation}
H_r=[\gamma_-\cos\phi-\rho(\gamma_+\cos^2\phi+\Gamma\sin^2\phi)]p_\rho+
p_\phi
[-\frac{\gamma_-\sin\phi}{\rho}+\frac{\sin(2\phi)}{2}(\gamma_+-\Gamma)]+\sqrt{p_\phi^2+p_\theta^2\cot^2\phi}
.\nonumber
\end{equation}
From this, we deduce the following lemma:
\begin{lemma}
(i)- The angle $\theta$ is a cyclic variable and $p_\theta$ is a
first integral (symmetry of revolution).\\
(ii)- For $\gamma_-=0$, using the coordinate $r=\ln \rho$, the
Hamiltonian takes the form:
\begin{equation}\label{eq32}
H_r=-(\gamma_+\cos^2\phi+\Gamma\sin^2\phi)p_r+\sin(2\phi)(\gamma_+-\Gamma)p_\phi+
\sqrt{p_\phi^2+p_\theta^2\cot^2\phi}.
\end{equation}
Hence, $r$ is an additional cyclic variable and $p_r$ is a first
integral. The system is thus Liouville integrable.
\end{lemma}
As a consequence, we can deduce two properties. First of all, the
$z$-axis is an axis of revolution and the state $q_0=^t(0,0,1)$ is
a pole. This means that by making a rotation around ($Oz$) of the
extremal synthesis for the 2D-system, we generate the extremal
synthesis for the 3D-system.

More generally, we have for $\gamma_-=0$ a system on the
\emph{two-sphere of revolution} described by Eqs. (\ref{eq31b})
and (\ref{eq31c}) coupled with the one dimensional system
(\ref{eq31a}) describing the evolution of the physical variable
$\rho$ corresponding to the purity of the system. Moreover, the
system is invariant for the transformation $\phi\mapsto \pi-\phi$
which is associated to a reflexional symmetry with respect to the
equator for the system (\ref{eq31}) restricted to the two-sphere
of revolution. This property is crucial in the analysis of the
integrable case.

If $\gamma_-\neq 0$ then the situation is more intricate. The
extremals solutions of order 0 satisfy the equations which are
singular for $\rho=0$:
\begin{eqnarray}\label{eq10}
\dot{\rho}&=&\gamma_-\cos\phi-\rho(\gamma_+\cos^2\phi+\Gamma\sin^2\phi)\nonumber\\
\dot{\phi}&=&-\frac{\gamma_-\sin\phi}{\rho}+\frac{\sin(2\phi)}{2}(\gamma_+-\Gamma)+\frac{p_\phi}{Q}\\
\dot{\theta} &=& \frac{p_\theta\cot^2\phi}{Q}\nonumber ,
\end{eqnarray}
and
\begin{eqnarray}
\dot{p}_\rho &=&(\gamma_+\cos^2\phi+\Gamma\sin^2\phi)p_\rho-\frac{\gamma_-\sin\phi}{\rho^2}p_\phi\nonumber\\
\dot{p}_\phi
&=&[\gamma_-\sin\phi+\rho(\Gamma-\gamma_+)\sin(2\phi)]p_\rho-[-\frac{1}{\rho}\cos\phi\gamma_-+
(\gamma_+-\Gamma)\cos(2\phi)]p_\phi+\frac{p_\theta^2\cos\phi}{Q\sin^3\phi} \nonumber\\
\dot{p_\theta} &=& 0\nonumber ,
\end{eqnarray}
where $Q=\sqrt{p_\theta^2\cot^2\phi+p_\phi^2}$.
\subsection{Regularity analysis}
The smooth extremal curves solutions of $\vec{H}_r$ are not the
only extremals because more complicated behaviors are due to the
existence of the switching surface $\Sigma$: $H_1=H_2=0$. Hence,
in order to get singularity results, we must analyze the possible
connections of two smooth extremals crossing $\Sigma$ to generate
a piecewise smooth extremal. This can also generate complex
singularities of the Fuller type, where the switching times
accumulate. In our problem, the situation is less complex because
of the symmetry of revolution. The aim of this section is to make
the singularity analysis of the extremals near $\Sigma$.

The structure of optimal trajectories is described by the
following proposition.
\begin{proposition}
Every optimal trajectory is:
\begin{itemize}
\item Either an extremal trajectory with $p_\theta=0$ contained in
a meridian plane and time-optimal solution of the 2D-system, where
$u=(u_1,0)$.
\item Or subarcs solutions of $\vec{H}_r$, where
$p_\theta\neq 0$ with possible connections in the equator plane
for which $\phi=\pi/2$.
\end{itemize}
\end{proposition}
\begin{proof}
The first assertion is clear. If $p_\theta=0$ then extremals are
such that $\dot{\theta}=0$ and up to a rotation around the
$z$-axis, they correspond to solutions of the 2D-system. The
switching surface $\Sigma$ is defined by:
$p_\theta\cot\phi=p_\phi=0$. We cannot connect an extremal with
$p_\theta\neq 0$ to an extremal where $p_\theta=0$ since at the
connection the adjoint vector has to be continuous. Hence, the
only remaining possibility is to connect subarcs of $\vec{H}_r$
with $p_\theta\neq 0$ at a point of $\Sigma$ leading to the
conditions $p_\phi=0$ and $\phi=\pi/2$.
\end{proof}
Further work is necessary to analyze the behaviors of such
extremals near $\Sigma$.\\
\textbf{Normal form}: A first step in the analysis is to construct
a normal form. Taking the system in spherical coordinates and
setting $\psi=\pi/2-\phi$, the approximation is:
\begin{eqnarray}
\dot{\rho}&=& \gamma_-\psi-\rho[\Gamma+(\gamma_+-\Gamma)\psi^2]\nonumber\\
\dot{\psi}&=& \frac{\gamma_-}{\rho}(1-\psi^2/2)-\psi(\gamma_+-\Gamma)-v_2\nonumber\\
\dot{\theta} &=& -\psi v_1\nonumber ,
\end{eqnarray}
with the corresponding Hamiltonian:
\begin{equation}
H_r=p_\rho[\gamma_-\psi-\rho(\Gamma+(\gamma_+-\Gamma)\psi^2)]+p_\psi[\frac{\gamma_-}{\rho}(1-\psi^2/2)-\psi(\gamma_+-\Gamma)]+
\sqrt{p_\psi^2+p_\theta^2\psi^2}.
\end{equation}
\begin{proposition}\label{prop2}
Near $\psi=0$, $p_\psi=0$, we have two distinct cases for optimal
trajectories:
\begin{itemize}
\item If $\gamma_-=0$, for the 2D-system, the line $\psi=0$ is a
singular trajectory with admissible zero control if
$\gamma_+-\Gamma\neq 0$. It is slow if $(\gamma_+-\Gamma)>0$ and
fast if $(\gamma_+-\Gamma)<0$. Hence, for this system, we get only
extremal trajectories through $\Sigma$ in the case
$(\gamma_+-\Gamma)<0$, where $\psi$ is of order $t$ and $p_\psi$
of order $t^2$. They are the only non-smooth optimal trajectories
passing through $\Sigma$.
\item If $\gamma_-\neq 0$, for the 2D-system, the set
$\psi=p_\psi=0$ becomes a set of ordinary switching points where
$\psi$ and $p_\psi$ are of order $t$. Moreover, connections for
extremals of $\vec{H}_r$ are eventually possible, depending upon
the set of parameters and initial conditions.
\end{itemize}
\end{proposition}
\begin{proof}
For the normal form, the adjoint system is:
\begin{eqnarray}
\dot{p}_\rho &=& p_\rho(\Gamma+(\gamma_+-\Gamma)\psi^2)+\frac{p_\psi}{\rho^2}\gamma_-(1-\frac{\psi^2}{2})\nonumber\\
\dot{p}_\psi &=& -p_\rho(\gamma_--2\psi\rho(\gamma_+-\Gamma))+p_\psi(\frac{\gamma_-\psi}{\rho}+(\gamma_+-\Gamma))+v_1p_\theta.\nonumber\\
\end{eqnarray}
In order to make the evaluation of smooth arcs reaching or
departing from $\Sigma$, the technique is simple: a solution of
the form $\psi(t)=at+o(t)$, $p_\psi(t)=bt+o(t)$ is plug in the
equations to determine the coefficients. From the equations, we
observe that the contacts with $\Sigma$ differ in the case
$\gamma_-=0$ from the case $\gamma_-\neq 0$ that we discuss
separately.

First of all, we consider the case $\gamma_-=0$; $p_\theta=0$,
$\psi=0$ is an admissible singular direction (with zero control)
which can be slow if $(\gamma_+-\Gamma)>0$ or fast if
$(\gamma_+-\Gamma)<0$. In the first case, there is no admissible
extremal through $\Sigma$ while it is possible if
$\gamma_+-\Gamma<0$. If we compute the different orders, we have
that $\psi$ is of order $t$, $p_\psi$ is of order $t^2$ while
$p_\rho$ has to be non zero if $p_\theta=0$. If we consider
extremals with $p_\theta\neq 0$, we can conclude with the orders
alone. Indeed the Hamiltonian is $H_r=\varepsilon$,
$\varepsilon=0,1$ and in both cases, we have:
\begin{equation}
-p_\rho\rho(\gamma_+-\Gamma)\psi^2-p_\psi\psi(\gamma_+-\Gamma)+\sqrt{p_\psi^2+p_\theta^2\psi^2}=0.\nonumber
\end{equation}
The conclusion using orders is then straightforward. For instance,
if $\psi$ and $p_\psi$ are of order one, this gives
$p_\psi=p_\theta\psi=0$ which is impossible. The other cases are
similar.

In the case $\gamma_-\neq 0$, the analysis is more intricate and
we must analyze the equations. We introduce the Hamiltonians:
\begin{equation}
H_1=-p_\theta\psi,~H_2=p_\psi .\nonumber
\end{equation}
Differentiating $H_1$ and $H_2$ with respect to $t$, one gets:
\begin{eqnarray}
\dot{H}_1 &=& \{H_1,H_0\}+v_2\{H_1,H_2\}\nonumber\\
\dot{H}_2 &=& \{H_2,H_0\}+v_1\{H_2,H_1\}\nonumber
\end{eqnarray}
and at a point of $\Sigma$, we obtain the relations:
\begin{equation}
\dot{H}_1=
-p_\theta(\gamma_--v_2),~\dot{H_2}=\gamma_-p_\rho-v_1p_\theta
.\nonumber
\end{equation}
In order to analyze the singularity, we use a polar blowing up:
\begin{equation}
H_1=r\cos\alpha,~H_2=r\sin\alpha ,\nonumber
\end{equation}
and we get:
\begin{eqnarray}
\dot{r}&=& \gamma_-[-\frac{p_\theta\cos\alpha}{\rho}+p_\rho\sin\alpha]\nonumber\\
\dot{\alpha}&=&
\frac{1}{r}[\gamma_-p_\rho\cos\alpha+\frac{p_\theta\gamma_-\sin\alpha}{\rho}-p_\theta].\nonumber
\end{eqnarray}
Hence, the extremals crossing $\Sigma$ are given by solving
$\dot{\alpha}=0$, while the sign of $\dot{r}$ is given by the
first equation above.

Depending upon the parameters and the initial conditions on
$(p_\rho,\rho)$, the equation $\dot{\alpha}=0$ can have at most
two distinct solutions on $(0,2\pi)$, while in the case
$p_\theta=0$, we get an ordinary switching point for the
single-input system. The assertion \ref{prop2} is proved.
\end{proof}
\subsection{Geometric analysis and numerical solution}
We first analyze the integrable case $\gamma_-=0$. We only present
a summary of the result of \cite{bonnardsugny} in order to be
generalized to the case $\gamma_-\neq 0$.
\subsubsection{The case $\gamma_-=0$}
The system (\ref{eq32}) is associated to a system on the
two-sphere of revolution of the form:
\begin{equation}
\dot{q}=G_0(q)+\sum_{i=1}^2u_iG_i(q).\nonumber
\end{equation}
It defines a Zermelo navigation problem on the two-sphere of
revolution where the drift term $G_0$ represents the current:
\begin{equation}
G_0=\frac{\sin(2\phi)}{2}(\gamma_+-\Gamma)\frac{\partial}{\partial
\phi},
\end{equation}
and $G_1=\frac{\partial}{\partial \phi}$,
$G_2=-\cot\phi\frac{\partial}{\partial \theta}$ form a frame for
the metric $g=d\phi^2+\tan^2\phi d\theta^2$ which is singular at
the equator $\phi=\pi/2$. The drift can be compensated by a
feedback with $|u|<1$ if $|\gamma_+-\Gamma|<2$. This leads to the
following discussion.\\

\begin{figure}
\centering
\includegraphics[width=0.4\textwidth]{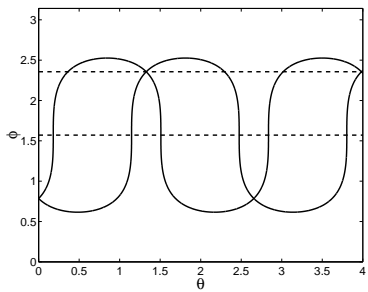}
\caption{\label{fig100} Extremal trajectories for $\Gamma=2.5$,
$\gamma_+=2$ and $\gamma_-=0$. Other parameters are taken to be
$p_\phi(0)=-1$ and 2.33, $\phi(0)=\pi/4$, $p_\rho=1$ and
$p_\theta=2$. Dashed lines represent the equator and the antipodal
parallel located at $\phi=3\pi/4$.}
\end{figure}

\begin{figure}
\centering
\includegraphics[width=0.4\textwidth]{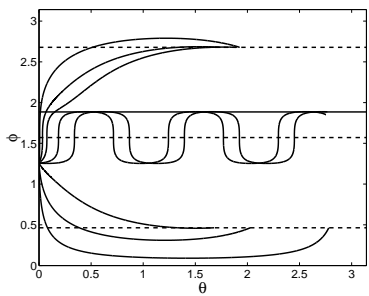}
\caption{\label{fig101} Extremal trajectories for $\Gamma=4.5$,
$\gamma_+=2$ and $\gamma_-=0$. Dashed lines represent the equator
and the locus of the fixed points of the dynamics. The solid line
corresponds to the antipodal parallel. Numerical values of the
parameters are taken to be $\phi(0)=2\pi/5$, $p_\theta=8$ and
$p_\rho(0)=0.25$. The different initial values of $p_\phi$ are
-50, -10, 0, 2.637, 3, 5, 10 and 50.}
\end{figure}

\textbf{Case $|\gamma_+-\Gamma|<2$}:  In this case, the system
reduced to the two-sphere defines a Finsler geometric for which
the extremals are a deformation of the extremals of
$g=d\phi^2+\tan^2\phi d\theta^2$. The main problem properties are
described in the next proposition.
\begin{proposition}\label{prop32}
If for fixed $(p_r,p_\theta)$, the level set of $H_r=\varepsilon$
($\varepsilon=0,1$) is compact without singular point and has a
central symmetry with respect to $(\phi=\pi/2,p_\phi=0)$ then it
contains a periodic trajectory $(\phi,p_\phi)$ of period $T$ and
if $p_\phi^\pm(0)$ are distinct, we have two distinct extremal
curves $q^+(t)$, $q^-(t)$ starting from the same point and
intersecting with the same length $T/2$ at a point such that
$\phi(T/2)=\pi-\phi(0)$ (see Fig. \ref{fig100}).
\end{proposition}
\textbf{Case $|\gamma_+-\Gamma|>2$}: We have two types of
extremals characterized by their projection on the two-sphere:
those occurring in a band near the equator and described by
proposition \ref{prop32} and those crossing a band near
$\phi=\pi/4$ and with asymptotic properties of proposition
\ref{prop33}:
\begin{proposition}\label{prop33}
If $|\Gamma-\gamma_+|\geq 2$ then we have extremal trajectories
such that $\dot{\phi}\to 0$, $|p_\phi|\to +\infty$ when $t\to
+\infty$ while $\dot{\theta}\to 0$.
\end{proposition}
Both behaviors are represented on Fig. \ref{fig101}.
\subsubsection{The case $\gamma_-\neq 0$}
\begin{figure}[ht]
\centering
\includegraphics[width=0.4\textwidth]{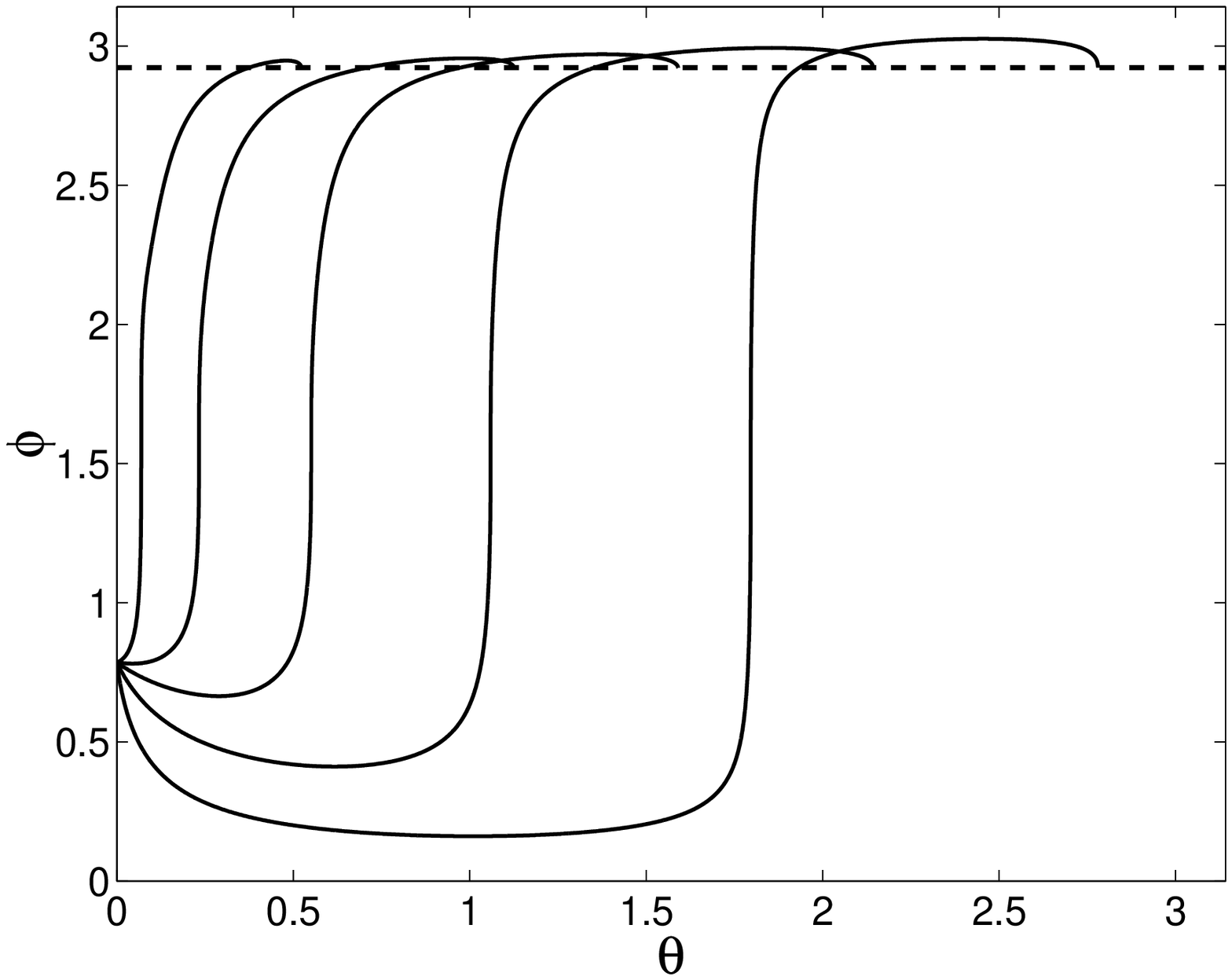}
\includegraphics[width=0.4\textwidth]{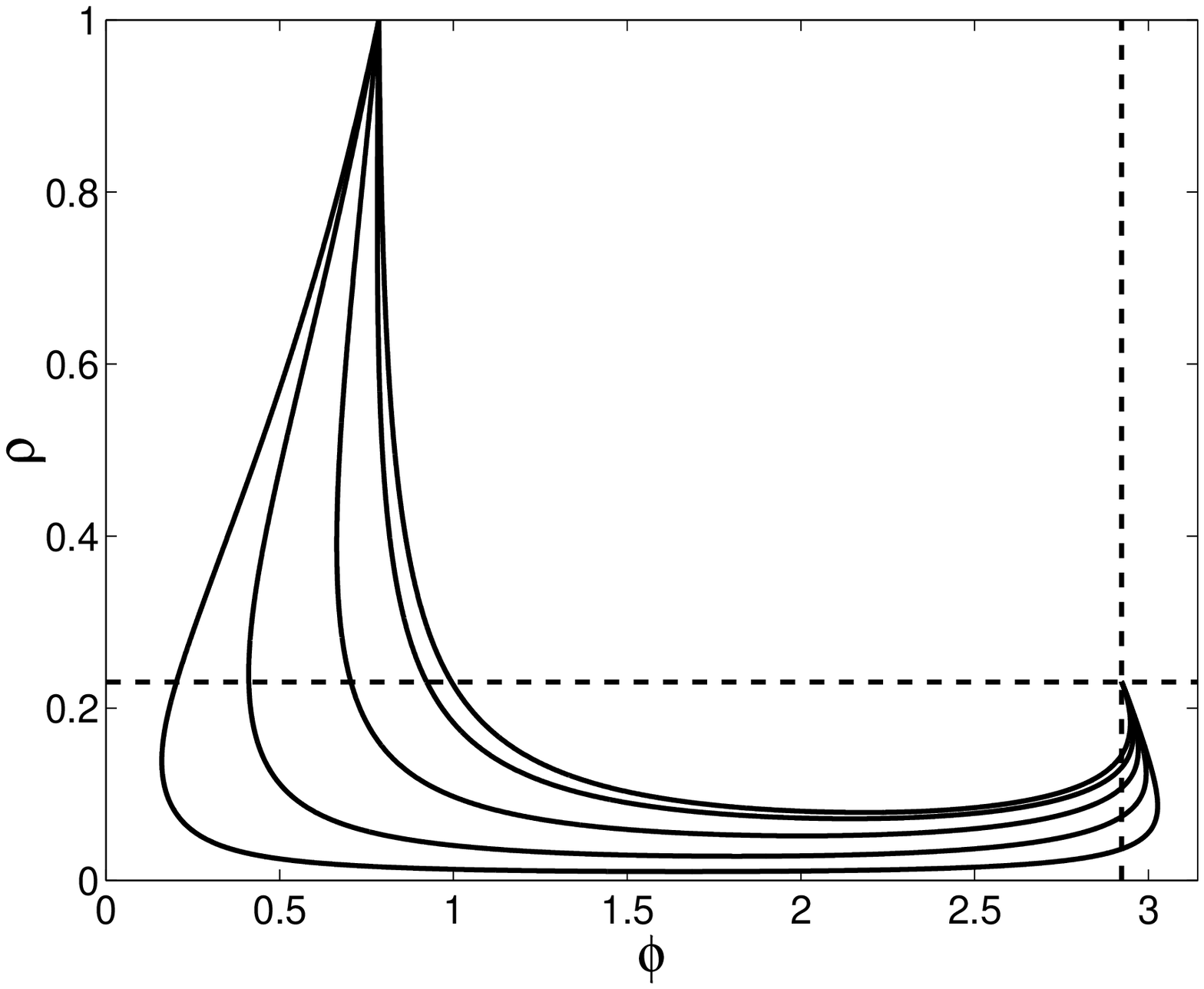}
\caption{\label{fig10} Extremal trajectories for $\Gamma=4.5$,
$\gamma_+=2$ and $\gamma_-=-0.5$. The equations of the dashed
lines are $\phi=\pi-\arctan(1/\Gamma)$ and
$\rho=|\gamma_-|\sqrt{1+\Gamma^2}/(1+\gamma_+\Gamma)$ (see the
text). Numerical values of the parameters are taken to be
$\phi(0)=\pi/4$, $p_\theta=2$, $p_\rho(0)=0.1$ and $\rho(0)=1$.
$p_\phi(0)$ is successively equal to -10, -2.5, 0, 2.5 and 10 for
the different extremals.}
\end{figure}

\begin{figure}[ht]
\centering
\includegraphics[width=0.4\textwidth]{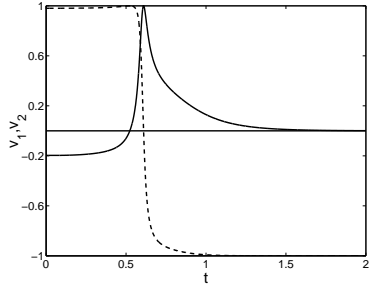}
\caption{\label{fig200} Plot of the optimal control fields $v_1$
(solid line) and $v_2$ (dashed line) as a function of time $t$ for
the extremal trajectory of Fig. \ref{fig10} with $p_\phi(0)=5$.
The equation of the horizontal solid line is $v=0$.}
\end{figure}

\begin{figure}[ht]
\centering
\includegraphics[width=0.4\textwidth]{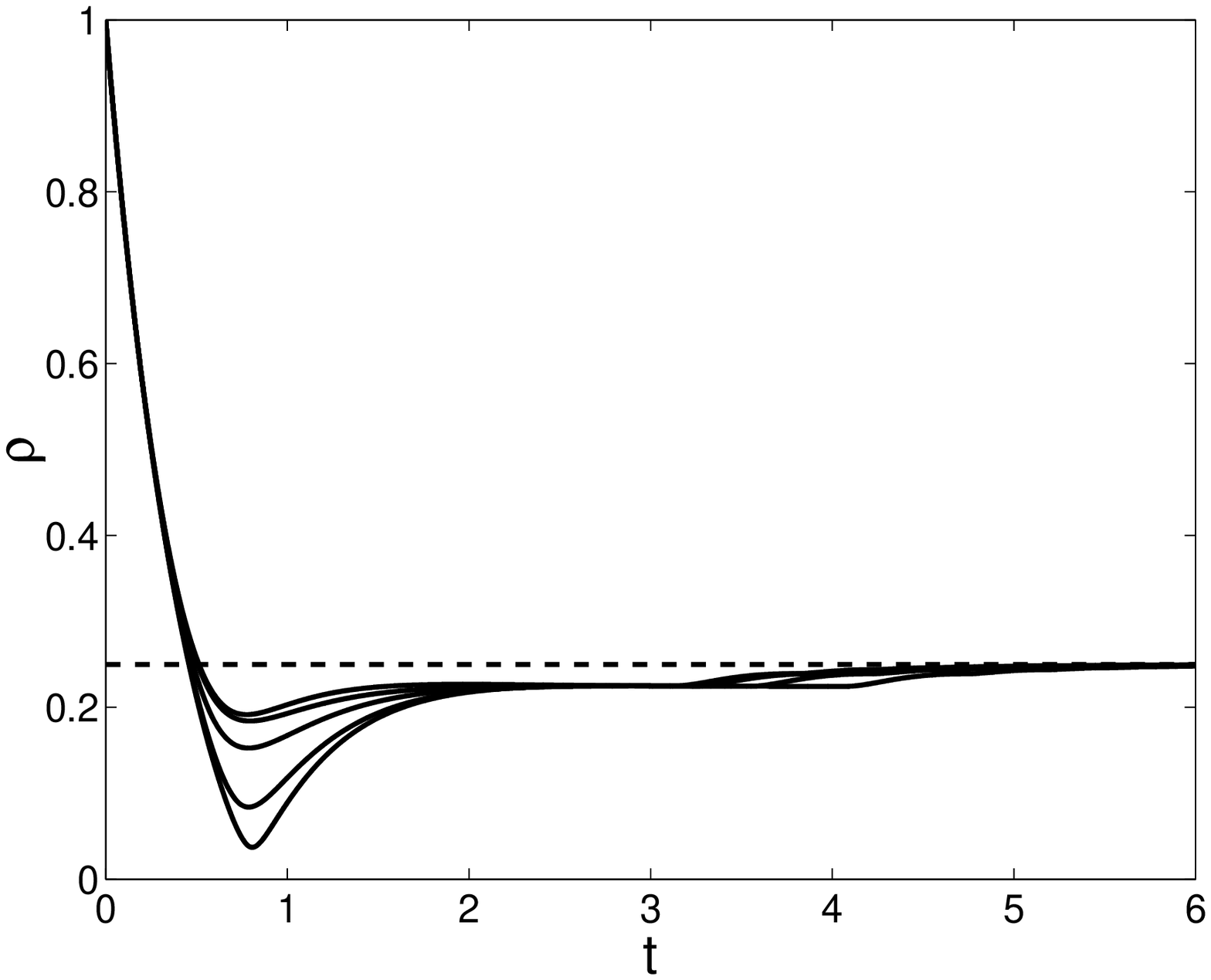}
\includegraphics[width=0.4\textwidth]{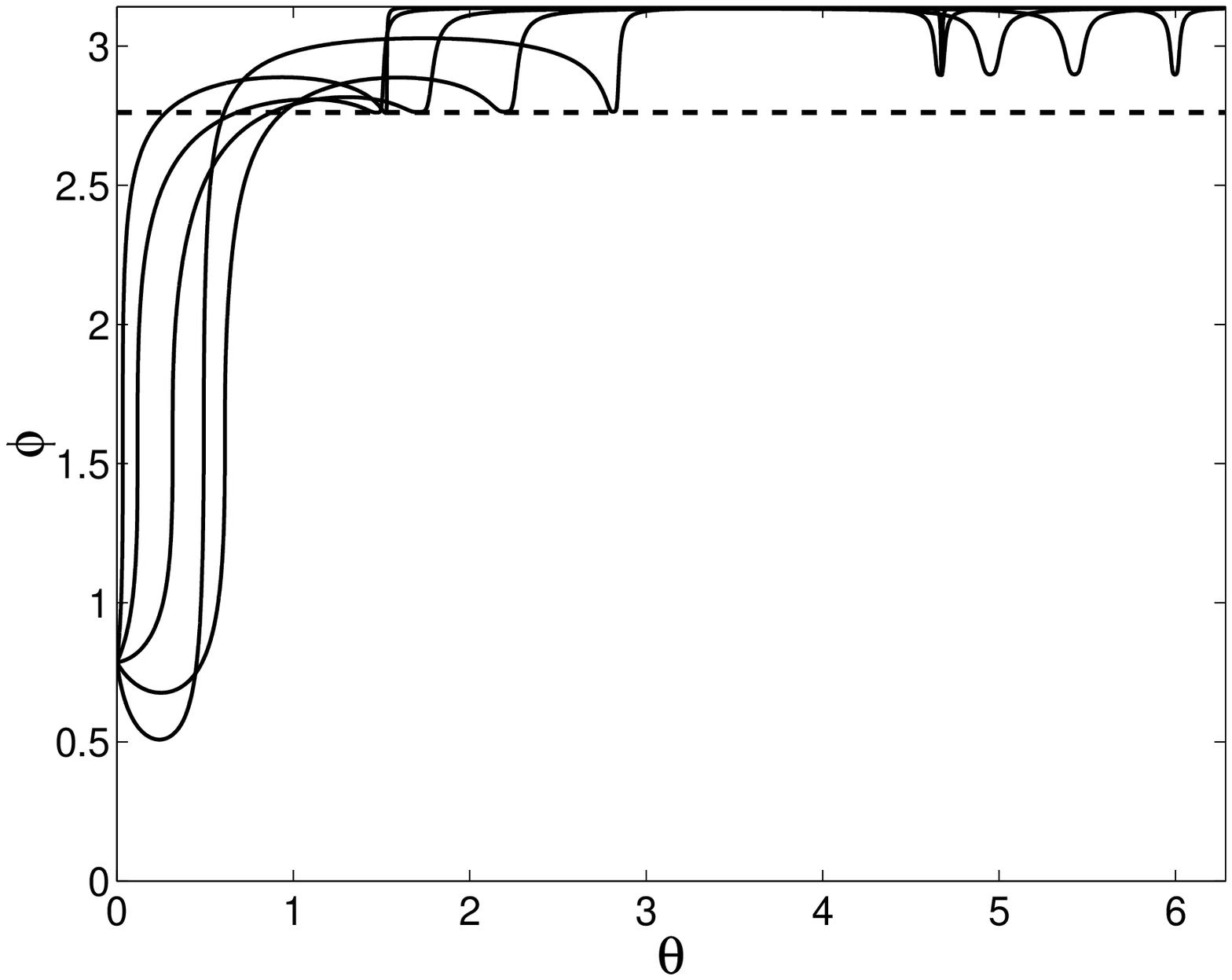}
\caption{\label{fig11} Same as Fig. \ref{fig10} but for
$\Gamma=2.5$. The equation of the dashed line is
$\rho=|\gamma_-|/\gamma_+$.}
\end{figure}

\begin{figure}[ht]
\centering
\includegraphics[width=0.4\textwidth]{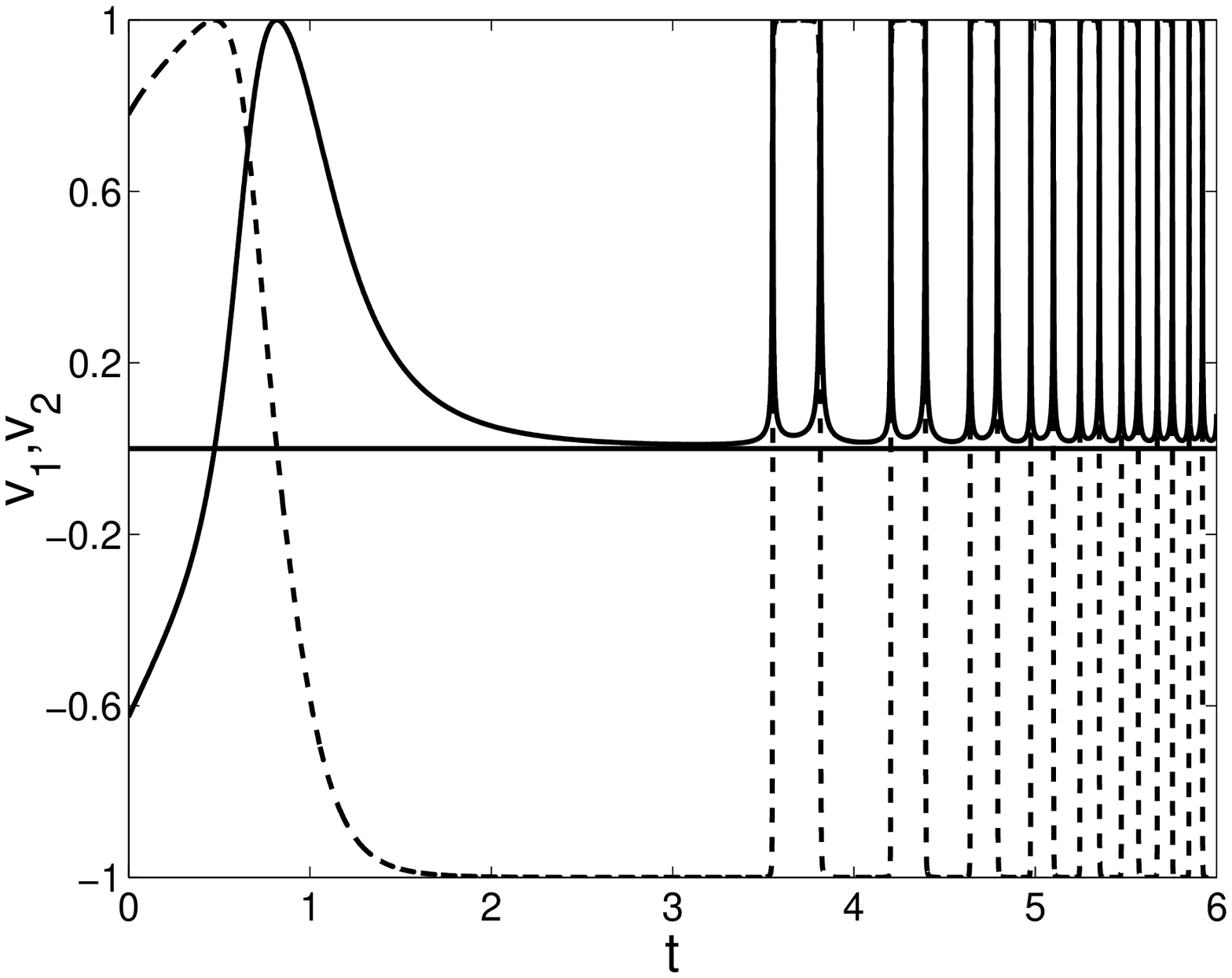}
\includegraphics[width=0.4\textwidth]{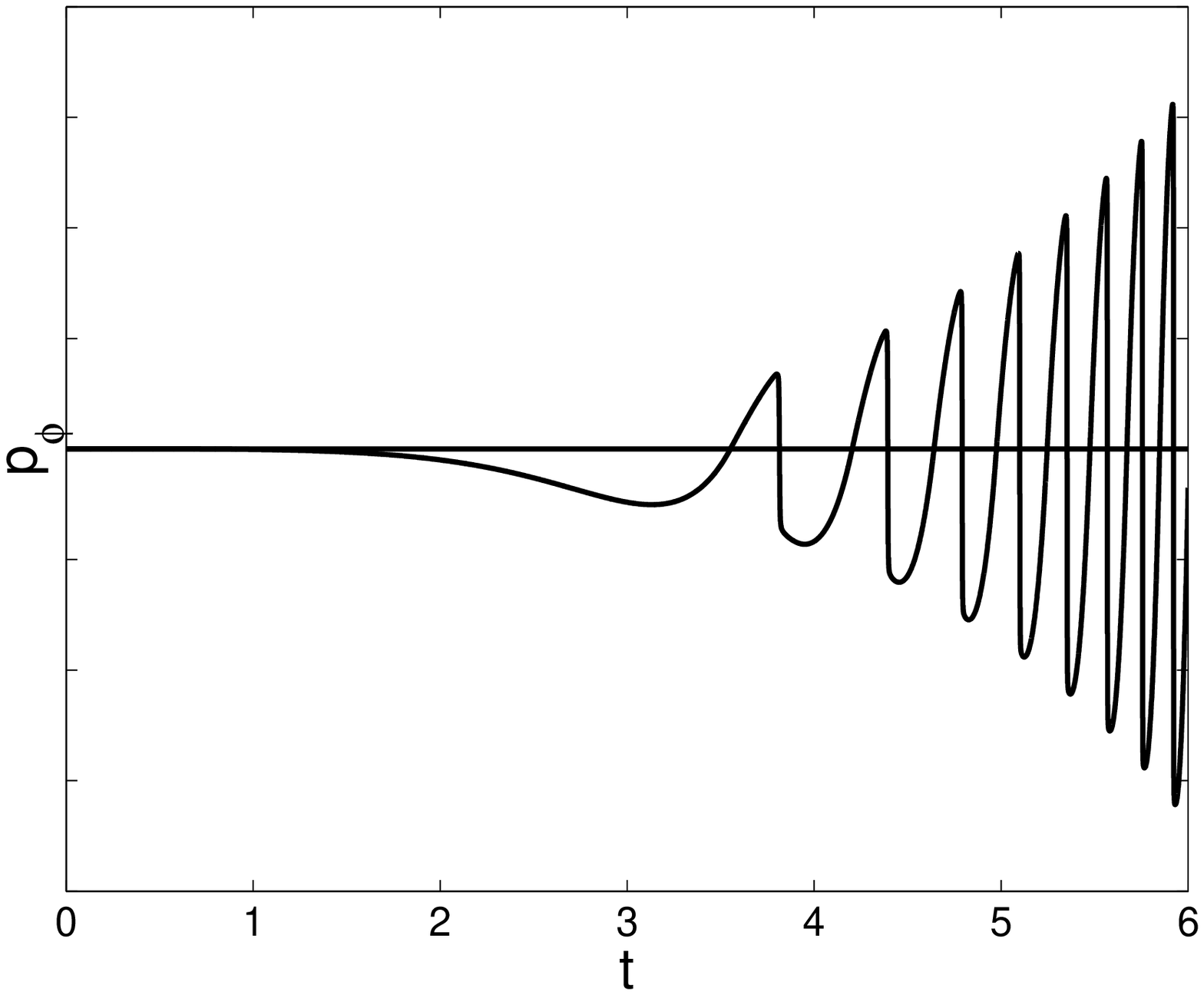}
\caption{\label{fig201} (top) Same as Fig. \ref{fig200} but for
the extremal of Fig. \ref{fig11} with $p_\phi(0)=2.5$. (bottom)
Evolution of $p_\phi$ for the same extremal as a function of $t$.}
\end{figure}

\begin{figure}[ht]
\centering
\includegraphics[width=0.4\textwidth]{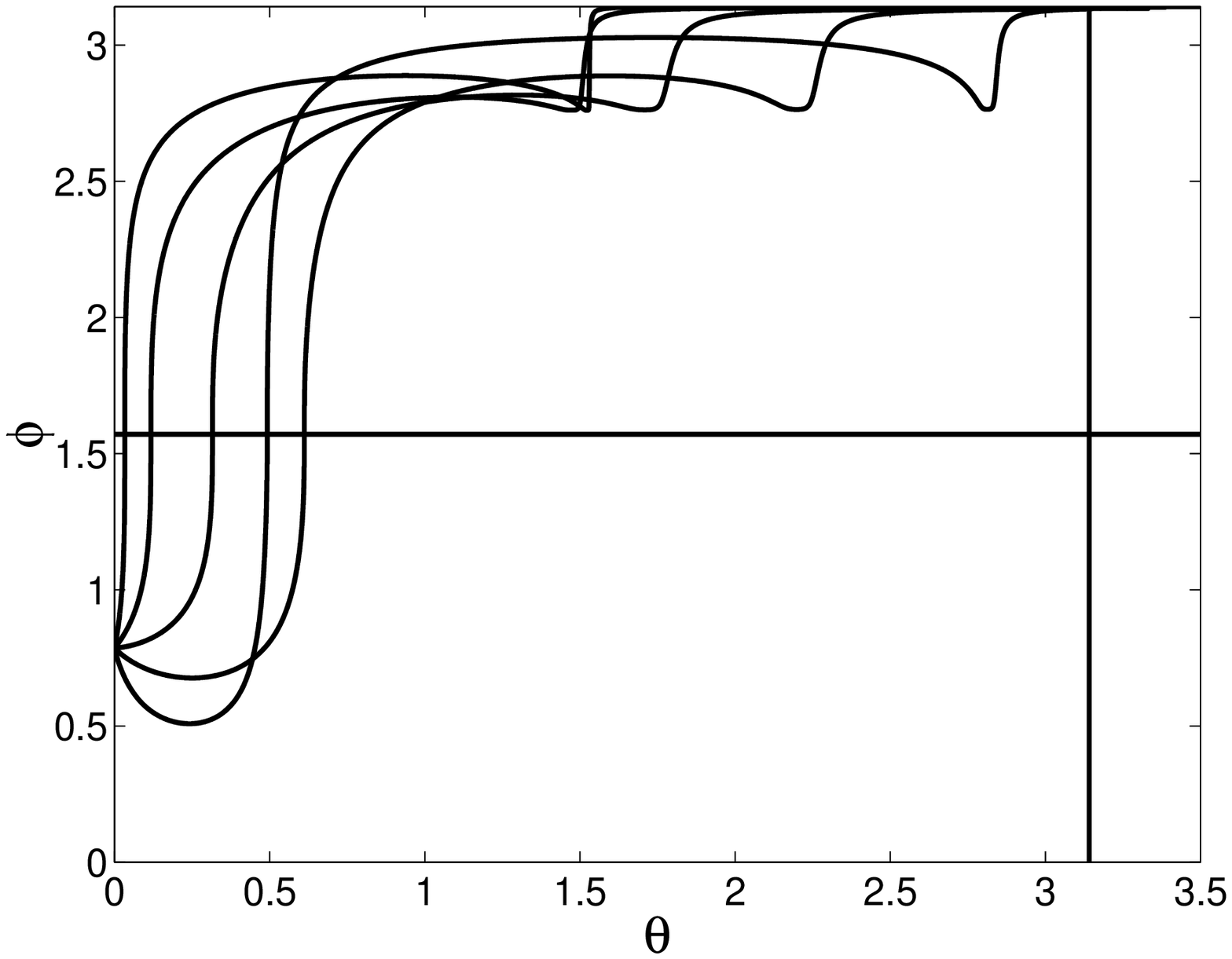}
\caption{\label{fig203} Plot of the extremals of Fig. \ref{fig11}
up to the first conjugate point. The coordinates $\theta$ of the
conjugate points are respectively 3.149, 3.116, 3.332, 3.386 and
3.535 for $p_\phi(0)$ equal to -10, -2.5, 0, 2.5 and 10. The
equations of the horizontal and vertical solid lines are
respectively $\phi=\pi/2$ and $\theta=\pi$.}
\end{figure}

We present numerical results about the behavior of extremal
solutions of order 0 and conjugate point analysis.\\ \\
\textbf{Extremal trajectories}:\\
We begin by analyzing the structure of extremal trajectories. The
description is based on a direct integration of the system
(\ref{eq10}). We observe two different asymptotic behaviors
corresponding to stationary points of the dynamics which are
described by the following results.
\begin{proposition}\label{propa}
In the case denoted (a) where $|p_\phi(t)|\to +\infty$ when $t\to
+\infty$, the asymptotic stationary points
$(\rho_f,\phi_f,\theta_f)$ of the dynamics are given by
$\rho_f=|\gamma_-|\sqrt{1+\Gamma^2}/(1+\gamma_+\Gamma)$ and
$\phi_f=\arctan(1/\Gamma)$ if $\gamma_->0$ or
$\phi_f=\pi-\arctan(1/\Gamma)$ if $\gamma_-<0$.
\end{proposition}
\begin{proof}
We assume that $|p_\phi(t)|\to +\infty$ as $t\to +\infty$ and that
$\cot(\phi)$ remains finite in this limit. One deduces from the
system (\ref{eq10}) that $(\rho_f,\phi_f)$ satisfy the following
equations:
\begin{eqnarray}
& &\gamma_-\cos\phi_f=\rho_f(\gamma_+\cos^2\phi_f+\Gamma\sin^2\phi_f)\nonumber\\
&
&\frac{\gamma_-\sin\phi_f}{\rho_f}=(\gamma_+-\Gamma)\cos\phi_f\sin\phi_f+\varepsilon
,\nonumber
\end{eqnarray}
where $\varepsilon=\pm 1$ according to the sign of $p_\phi$. The
quotient of the two equations leads to
\begin{equation}
(\gamma_+-\Gamma)\cos\phi_f\sin\phi_f+\varepsilon=\tan\phi_f(\gamma_+\cos^2\phi_f+\Gamma\sin^2\phi_f)\nonumber
\end{equation}
which simplifies into
\begin{equation}
\tan\phi_f=\frac{\varepsilon}{\Gamma}.\nonumber
\end{equation}
Using the fact that $\phi_f\in ]0,\pi[$ and
$\gamma_-\cos\phi_f\geq 0$, one arrives to
$\phi_f=\arctan(1/\Gamma)$ if $\gamma_->0$ and
$\phi_f=\pi-\arctan(1/\Gamma)$ if $\gamma_-<0$. From the equation
\begin{equation}
\gamma_-\cos\phi_f=\rho_f(\gamma_+\cos^2\phi_f+\Gamma\sin^2\phi_f)\nonumber,
\end{equation}
one finally obtains that
\begin{equation}
\rho_f=\frac{\gamma_-\sqrt{1+\Gamma^2}}{1+\gamma_+\Gamma}\nonumber
.
\end{equation}
\end{proof}
\begin{proposition}\label{propb}
In the case denoted (b) where $\lim_{t\to +\infty}\phi(t)=0$ or
$\pi$, the asymptotic limit of the dynamics is characterized by
$\rho_f=|\gamma_-|/\gamma_+$ and $\phi_f=0$ if $\gamma_->0$ or
$\phi_f=\pi$ if $\gamma_-<0$.
\end{proposition}
\begin{proof}
Using the relation
\begin{equation}
\gamma_-\cos\phi_f=\rho_f(\gamma_+\cos^2\phi_f+\Gamma\sin^2\phi_f),\nonumber
\end{equation}
one deduces that $\gamma_-\cos\phi_f\geq 0$ and that
$\rho_f=|\gamma_-|/\gamma_+$ if $\phi_f=0$ or $\pi$.
\end{proof}
We have numerically checked that if $|\Gamma-\gamma_+|>2$ then
only the case (a) is encountered whereas if $|\Gamma-\gamma_+|<2$,
the extremals are described by cases (a) and (b). One
particularity of the case (a) is the fact that the limit of the
dynamics only depends on $\Gamma$ and on the sign of $\gamma_-$
and not on $\phi(0)$ or $\gamma_+$. The structure of the extremals
is also simple in case (b) since the limit of $\phi$ is 0 or $\pi$
independently of the values of $\Gamma$, $\gamma_+$ or $\gamma_-$.
The different behaviors of the extremals are illustrated in Fig.
\ref{fig10} for the case $|\Gamma-\gamma_+|>2$ and in Fig.
\ref{fig11} for the case $|\Gamma-\gamma_+|<2$. The corresponding
optimal control fields $v_1$ and $v_2$ are represented in Fig.
\ref{fig200} for the case (a) and in Fig \ref{fig201} for the case
(b). In Fig. \ref{fig200}, note that the control $v_1$ tends to 0
whereas $v_2$ is close to -1 for $t$ sufficiently large. This is
due to the fact that $|p_\phi|\to +\infty$ when $t\to +\infty$ and
can be easily checked from the definition of $v_1$ and $v_2$. We
observe a similar behavior for the case (b) in Fig. \ref{fig201}.
The control field $v_1$ acquires here a bang-bang structure which
is related to the unbounded and oscillatory behavior of
$p_\phi(t)$ (see Fig.
\ref{fig201}).\\ \\
\textbf{Conjugate points}:\\
The Cotcot code is used to evaluate the conjugate points. This
occurs only in case (b) and the numerical simulations give that
the first conjugate points appear before an uniform number of
oscillations of the $\phi$ variable. This phenomenon is
represented on Fig. \ref{fig203}. Cutting the trajectory at the
first conjugate point avoids such a behavior. Note that due to the
symmetry of revolution, the global optimality is lost for
$\theta\leq \pi$.
\section{Physical conclusions}
We give some qualitative conclusions on the time-optimal control
of two-level dissipative systems. The discussion concerns the role
of dissipation which can be beneficial or not for the dynamics and
the robustness with respect to dissipative parameters of the
optimal control.

The dissipation effect is well summarized by Fig. \ref{fig3}. In
this case $\Gamma>\gamma_+$ and one sees that as long as the
purity of the state decreases (for $0\leq z\leq 1$), it is
advantageous to use a control field, the dissipation being
undesirable. On the contrary, when the purity starts increasing
(for $-\gamma_-/\gamma_+\leq z\leq 0$) then the dissipation alone
becomes more efficient and its role positive. The quickest way to
accelerate the purification of the state consists in letting the
dissipation acts. This constitutes a non-intuitive physical
conclusion which, however, crucially depends on the respective
values of $\Gamma$ and $\gamma_+$. For instance, if
$\gamma_+>\Gamma$ then all the preceding conclusions are modified.

The robustness of the optimal control with respect to dissipative
parameters is illustrated by the double-input control. We give
different examples. If $\gamma_-=0$ then the integrability of the
Hamiltonian and the geometrical properties of the extremals are
preserved when $|\Gamma-\gamma_+|<2$. If $\gamma_-\neq 0$ then the
asymptotic behavior of the extremals slightly depends on the
parameters $\Gamma$, $\gamma_+$ and $\gamma_-$ (see Propositions
\ref{propa} and \ref{propb}). Fig. \ref{fig200} and \ref{fig201}
show that the extremal control fields have also asymptotic
behaviors independent of the dissipation. In case (a), the control
fields tend to a constant whereas a bang-bang structure appears in
case (b). This conclusion could be interesting for practical
applications where robustness with respect to physical parameters
and simple control fields are needed. In addition, due to the
simple structure of the time-optimal synthesis, shooting
techniques will be particularly efficient to determine the control
fields especially in case (a).
\section*{Acknowledgment}
We acknowledge support from the Agence Nationale de la recherche
(ANR CoMoc).
\bibliographystyle{IEEEtran}

\end{document}